\documentclass[journal=jctcce,manuscript=article]{achemso}

\usepackage{graphicx}
\usepackage{booktabs}
\usepackage{amsmath}
\usepackage{amssymb}
\usepackage[utf8]{inputenc}
\usepackage[T1]{fontenc}
\usepackage{xcolor}
\usepackage{hyperref}
\usepackage{float}
\usepackage{pdfpages}

\title{From Heuristics to Machine Learning: The Performance Ceiling for Single-Ion Magnets and Its Electronic Origin}

\author{Federico Zahariev}
\email{fzahari@iastate.edu}
\affiliation{Ames National Laboratory and Department of Chemistry, Iowa State University, Ames, Iowa 50011, USA}

\author{Regina Pereyra}
\affiliation{Ames National Laboratory and Department of Chemistry, Iowa State University, Ames, Iowa 50011, USA}

\author{Vassiliki-Alexandra Glezakou}
\email{glezakouva@ornl.gov}
\affiliation{Chemical Sciences Division, Oak Ridge National Laboratory, Oak Ridge, Tennessee 37830, USA}

\author{Durga Paudyal}
\affiliation{Department of Physics and Astronomy, University of Iowa, Iowa City, Iowa 52242, USA}
\email{durga.quantum@gmail.com}

\begin{document}

\begin{abstract}
The machine-learning (ML) community has been optimistic about its ability to speed up the discovery of single-ion magnets (SIMs) and other magnetic materials. However, does the information contained in their structure allow us to make predictions? In order to answer this question, we have taken a subset of 1215 lanthanide complexes from the SIMDAVIS 1.2.1 database and compared three increasing levels of structural information, such as: i) tabular descriptions at the level of individual coordination sites; ii) symmetry measures, which describe the deviation of the coordination polyhedron from ideal shapes; iii) a complete description of the 3D arrangement of atoms in space. All of the above cases converge to an accuracy near $\sim$76\%, which is just slightly higher than that of the one rule Predict SIM for Dy$^{3+}$, which was $\sim$71\%. We used multireference ab initio methods to explain why ML could not accurately predict every SIM, and then inspected the structural representation of the high-confidence failures. We found that all SIMs missed by geometric ML models were field-induced relaxers containing tunnelling-prone Kramers doublets that the geometry-based methods could not identify. On the other hand, many of the false positives identified in the study had multiple lanthanide centers or radicals and hence possessed properties that were influenced collectively rather than individually and thus could not be described using single-ion models. As a result, we conclude that both the electronic-structural and connectivity information required to identify SIMs (or other magnetic materials) may not be available via geometric methods alone. Nonetheless, our work demonstrates the utility of such geometric-based models: by restricting screening to those samples where the model's prediction confidence is very high, we can increase the overall accuracy of our prediction to $\sim$88\% while still retaining $\sim$48\% of the original dataset. Additionally, by examining the reasons for failure, we can suggest a practical strategy to go beyond the current limitations of geometric ML models for identifying SIMs. The strategy combines simple filters for nuclearity and the presence of radicals and/or unpaired electrons with ab initio calculated ligand-field-based descriptors.
\end{abstract}

\noindent\textbf{Keywords:} single-ion magnets; lanthanides; machine learning; SHAP; continuous symmetry measures; multireference ab initio; magnetic anisotropy

\section{Introduction}

Single-ion magnets (SIMs) are mononuclear lanthanide complexes that exhibit slow relaxation of magnetization arising from the crystal field splitting of the ground-state $J$ multiplet. In the years since Ishikawa and coworkers published the first SIM, [Pc$_2$Tb]$^-$,\cite{Ishikawa2003} barrier heights have increased to over 1500 cm$^{-1}$, and blocking temperatures are now approaching 80 K in dysprosium complexes.\cite{Goodwin2017,Guo2018} This progress was largely driven by chemists' ability to use their understanding of chemistry as a guide. As Dy$^{3+}$ has an oblate $4f$ electron density, strongly axial ligand fields stabilize states with large $|m_J|$ values, which generate the crystal field splitting that allows for SIM behavior.\cite{Rinehart2011} However, the accessible lanthanide chemical space is extremely large, and the relationships between molecular structure and magnetic relaxation are still poorly understood.

Machine learning (ML) offers a potential route to systematize the relationship between structure and SIM quality. Takiguchi and coworkers demonstrated that a 3D convolutional neural network could be trained to achieve 70\% accuracy using voxelized crystal structures from the Cambridge Structural Database.\cite{Takiguchi2024} Similarly, Frangoulis and coworkers used a combination of genetic algorithms and multireference ab initio calculations to develop computational candidate systems for Co(II) and Dy(III) molecular magnets.\cite{Frangoulis2025} Additionally, variational autoencoders were also employed to obtain estimates of crystal field parameters from susceptibility data.\cite{Escalera2018} While these studies demonstrate significant progress toward developing predictive tools for the design of lanthanide SIMs, there is still an important unanswered question: how much structural detail is required and sufficient to predict the quality of a SIM? This question matters for two reasons. First, practical SIM screening requires models that operate on information available before synthesis of materials or complexes---coordination number, ligand identity, ion choice---rather than crystal structures that require both synthesis and characterization. Second, understanding where prediction fails reveals what physics is missing from structural descriptors, guiding both experimental and computational efforts toward the most informative measurements.

We evaluated the SIMDAVIS 1.2.1 database\cite{Canon2025,Duan2022}---the largest curated collection of lanthanide SIM measurements, containing 1411 compounds representing 11 different chemical families---by systematically developing a hierarchy of predictive models. The progression included starting with a single-feature heuristic (``Is it Dy$^{3+}$?''), moving through a set of tabular coordination-level features, applying continuous symmetry measures (CSMs) from SHAPE analysis,\cite{SHAPE} and finally using a complete 3D representation of the crystal structure. In addition to assessing the incremental amount of structural information provided at each step, we applied SHAP (SHapley Additive exPlanations)\cite{Lundberg2017,Lundberg2020} to interpret the contribution of each feature to the prediction made for each compound, identifying both strong structural correlations and limitations associated with employing structure-based machine learning approaches for designing SIMs. First, we tested if the apparent ceiling exists across three increasingly rich structural representations. Next, we employed a combination of multireference ab initio calculations and structural analysis to assess the two most common error types. Lastly, we determined the range of structural richness in which screening based upon structure will remain reliable and what additional descriptors are required to exceed the ceiling.

\section{Dataset and Methods}

\subsection{The SIMDAVIS 1.2.1 Database}

SIMDAVIS 1.2.1 contains 1411 lanthanide complexes, collected from 451 separate scientific articles, classified by their magnetic behavior as determined from AC susceptibility measurements. A dataset of 1,215 compounds was used for classification, following the removal of 196 compounds with missing AC measurements. The dataset divides into two categories: 613 compounds exhibiting slow magnetic relaxation and 602 that do not. Of the 613 relaxing compounds, 320 feature frequency-dependent imaginary part of the complex susceptibility $\chi''$ and 293 feature a blocking temperature $T_{B3} > 2$~K. The database spans 10 lanthanide ions (Dy$^{3+}$: 601, Tb$^{3+}$: 207, Er$^{3+}$: 143, Yb$^{3+}$: 72, Ho$^{3+}$: 70, Gd$^{3+}$: 36, Nd$^{3+}$: 35, Tm$^{3+}$: 25, Pr$^{3+}$: 14, Ce$^{3+}$: 12) and 11 chemical families. The families are defined by the ligand environment following SIMDAVIS;\cite{Duan2022,Canon2025} short definitions and compound counts are given in Table~S1. We used the database version 1.2.1 (with release number r2025\_02\_18) for all statistical analyses that are presented here. Compounds published between 2003 and 2019 are represented, enabling temporal validation.

\subsection{Feature Engineering}

We constructed two feature sets from information available prior to synthesis. The basic feature set (30 features) comprised: coordination number (CN), number of ligands, concentration, anisotropy type (oblate/prolate), Kramers character, coordination element flags (oxygen, nitrogen, carbon), chemical family (one-hot encoded, 11 categories), and ion identity (one-hot encoded, 10 categories). The enhanced feature set (48 features) was generated by combining the continuous symmetry measure (CSM) from SHAPE analysis, the axial-distortion parameter, the CSM slope, the closest polyhedron type (encoded one-hot, ten categories), missingness indicators, and two engineered interactions (CSM$\times$oblate and axial distortion$\times$Kramers). CSM data was available for 727 of 1215 classification compounds (60\%); for the remaining compounds, missing values were imputed at the median and accompanied by missingness flags. The complete feature list and short definitions of the feature labels are given in Section~S2 and Table~S2 of the Supporting Information.

\subsection{Classification Framework}\label{sec:framework}

Gradient boosting was selected as our primary classifier, using 200 trees, a maximum depth of four, a learning rate of 0.05, and at least five samples per leaf. Gradient boosting was compared to logistic regression, random forest, decision trees, multilayer perceptrons, and an RBF-kernel support vector machine.\cite{Pedregosa2011} All features were scaled via min--max scaling between the range $[0, 1]$. Performance was evaluated by stratified 5-fold cross-validation, reporting accuracy, weighted F1, and AUC-ROC. Shapley values were computed via TreeExplainer for every compound in order to interpret every individual classification result.\cite{Lundberg2017} Paired McNemar tests\cite{McNemar1947,Dietterich1998} were employed to determine if there were statistically significant differences between each pair of classifiers. Bootstrap confidence intervals were obtained from 1,000 bootstrap resamples.

\subsection{Validation Protocols}

Beyond standard cross-validation, we employed four additional validation protocols: (i) Leave-one-family-out (LOFO): train on 10 families, predict the 11th. (ii) Temporal validation: train on compounds published before year $Y$, test on year $Y$, using actual publication dates from SIMDAVIS. (iii) Confidence-stratified evaluation: accuracy as a function of the model's prediction confidence ($|p-0.5|$). (iv) Cross-ion transfer: train on ion A, predict ion B. These protocols test fundamentally different aspects of generalization: to unseen chemistry, to future discoveries, to high-confidence screening, and to different electronic structures. Group-wise and time-ordered splits were used according to standard out-of-distribution validation practices.\cite{Meredig2018,Sheridan2013}

\section{Results and Discussion}

We first explore whether structural information alone supports prediction.

\subsection{How Much Do Structural Descriptors Buy?}

\begin{figure}[H]
\centering
\includegraphics[width=0.85\textwidth]{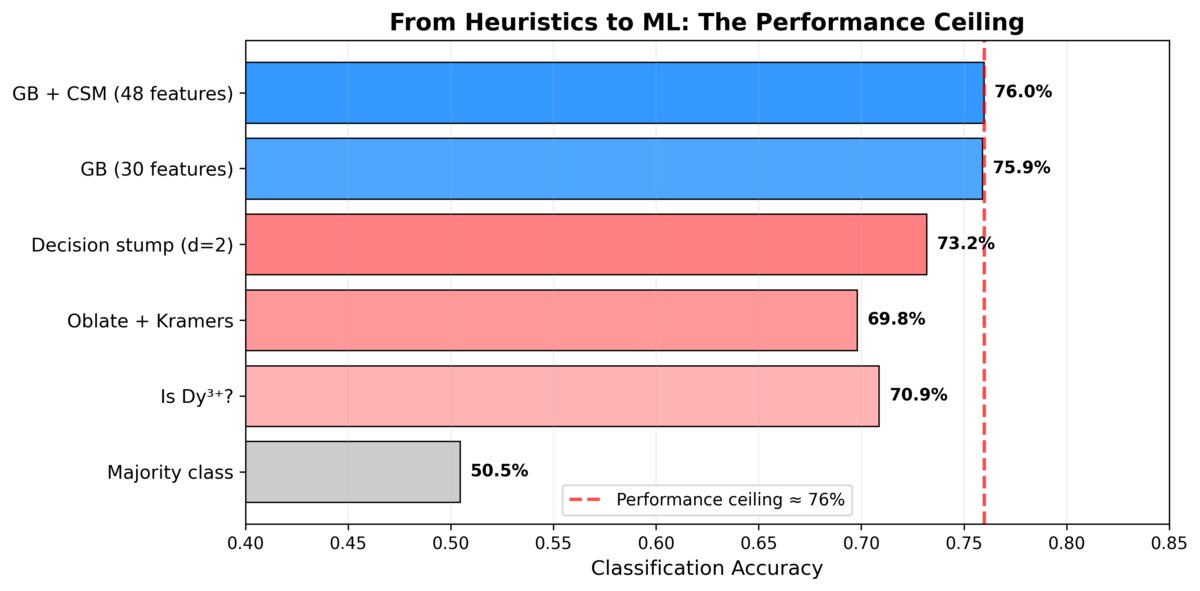}
\caption{How much do structural descriptors buy? Models are defined in Table~\ref{tab:baselines}. Accuracy converges near $\sim$76\% within the tabular descriptor space, with the ``Is it Dy$^{3+}$?'' heuristic already at 71\%.}
\label{fig:baselines}
\end{figure}

Table~\ref{tab:baselines} and Figure~\ref{fig:baselines} illustrate the comparison of predictive models of increasing representational complexity. The models are defined in the footnote of Table~\ref{tab:baselines}. The simplest meaningful heuristic---Predict SIM if Dy$^{3+}$---achieves 70.9\% accuracy, reflecting the dominance of Dy$^{3+}$ in the SIM literature and its genuinely superior electronic properties for axial magnetization. The depth-2 decision tree discovers the rule ``If Dy$^{3+}$ $\to$ SIM; else if LnPc$_2$ $\to$ SIM; else not SIM'' and achieves 73.2\%.

The full gradient boosting model with 30 basic features reaches 75.9\% accuracy (AUC = 0.836). Adding 18 CSM-based features yields 76.0\% accuracy (AUC = 0.826). The McNemar test confirms this difference is not statistically significant ($p = 1.0$); bootstrap analysis shows P(enhanced $>$ basic) = 0.54 with 95\% CI for $\Delta$accuracy of $[-0.016, +0.017]$.

We find it particularly interesting that the addition of 18 descriptors that relate directly to crystal field geometry does not impact the predictive performance of approximately 76\%. Within the space of coordination-level tabular descriptors tested here, accuracy saturates. The limiting factor is not geometric precision but rather information absent from any tabular representation: the full crystal field Hamiltonian, spin--orbit coupling magnitudes, Raman and QTM relaxation rates, and phonon-mediated pathways that depend on lattice dynamics inaccessible from single-molecule coordination descriptors.

As shown in the learning curve subsampling analysis (Figure S13), no plateau occurs when $N = 1000$: test accuracy increases from 62\% at $N = 100$ to 71\% at $N = 900$. Enhanced features show higher overfitting (larger train--test gap), supporting the interpretation that CSM features add complexity without adding generalizable information.

\begin{table}[H]
\centering
\caption{Classification accuracy from physics-based rules to enhanced ML. All models use 5-fold stratified cross-validation ($N = 1215$).}
\label{tab:baselines}
\begin{tabular}{lcccc}
\toprule
Model & $N_\text{feat}$ & Accuracy & AUC & $p$ vs basic \\
\midrule
Majority class & 0 & 0.505 & --- & --- \\
Is Dy$^{3+}$? & 1 & 0.709 & --- & --- \\
Oblate + Kramers & 2 & 0.698 & --- & --- \\
Decision stump (d=2) & 2 & 0.732 & --- & --- \\
Gradient Boost (basic) & 30 & 0.759 & 0.836 & ref. \\
Gradient Boost (enhanced) & 48 & 0.760 & 0.826 & 1.000 \\
\bottomrule
\end{tabular}
\begin{flushleft}\footnotesize
Majority class: every compound assigned to the more frequent class. Is Dy$^{3+}$?: SIM if the ion is Dy$^{3+}$. Oblate + Kramers: SIM if the ion is oblate and Kramers (Ce$^{3+}$, Nd$^{3+}$, Dy$^{3+}$). Decision stump (d=2): decision tree of depth two learned from the basic features. Gradient Boost (basic, enhanced): the gradient boosting classifier of Section~\ref{sec:framework} with the 30-feature basic or 48-feature enhanced set (``GB'' and ``GB + CSM'' in Figure~\ref{fig:baselines}). $N_\text{feat}$: number of input features. $p$ vs basic: McNemar $p$-value relative to Gradient Boost (basic).
\end{flushleft}
\end{table}

\subsection{SHAP Analysis: What Each Feature Contributes}

\begin{figure}[H]
\centering
\includegraphics[width=0.8\textwidth]{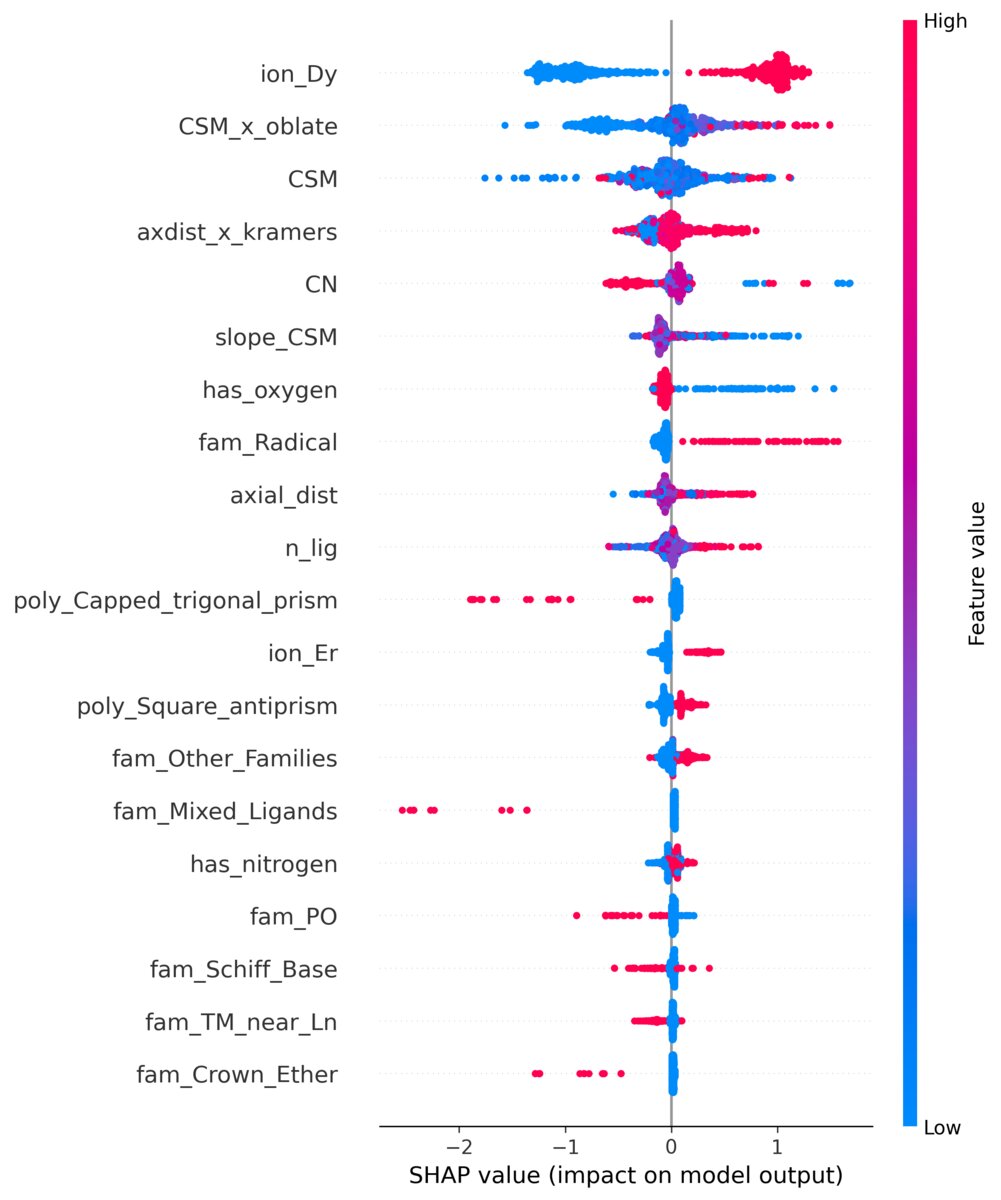}
\caption{SHAP beeswarm plot for the enhanced model (48 features, $N = 727$). CSM$\times$oblate is the second most important feature after Dy$^{3+}$ identity. Feature labels are defined in Table~S2.}
\label{fig:shap}
\end{figure}

SHAP expresses every prediction as a sum of feature contributions within a cooperative-game framework.\cite{Lundberg2017} As illustrated in Figure~\ref{fig:shap}, we provide the beeswarm summary for the enhanced model (48 features; 727 compounds with CSM data). The global hierarchy is: ion\_Dy (mean $|$SHAP$|$ = 0.95), CSM$\times$oblate (0.33), CSM (0.22), axial distortion$\times$Kramers (0.17), CN (0.16), slope\_CSM (0.15), has\_oxygen (0.14), fam\_Radical (0.14), axial\_distortion (0.11), $n_\text{lig}$ (0.09).

The CSM$\times$oblate interaction is the second most important feature by SHAP, meaning that the combination of symmetry deviation and anisotropy type carries more predictive power than either alone. However, this interaction is already implicitly captured by the combination of CN, family, and anisotropy features in the basic model---which is why adding CSM explicitly does not improve classification.

In the SHAP dependence plots (Figure S14), we demonstrate that the three CSM-related features influence predictions non-linearly and depend on context.

\subsection{Continuous Symmetry Measures and SIM Quality}

\begin{figure}[H]
\centering
\includegraphics[width=0.85\textwidth]{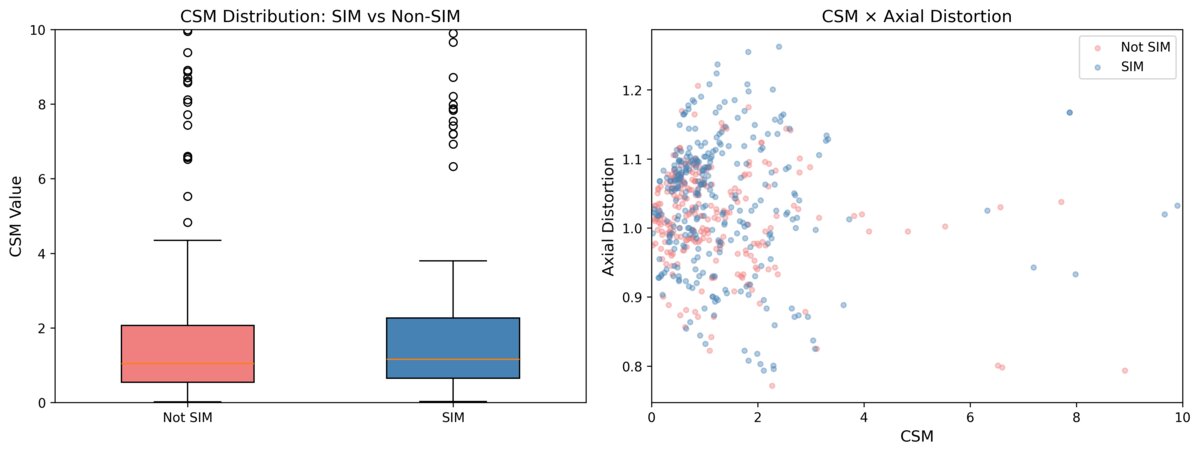}
\caption{CSM analysis. Left: CSM values for SIM vs non-SIM showing near-complete overlap. Right: CSM vs axial distortion, confirming no separable boundary exists.}
\label{fig:csm}
\end{figure}

In Figure~\ref{fig:csm}, we compare CSM values and success rates of SIMs across coordination polyhedra. The results are strikingly non-monotonic: square antiprism (CSM = 0.80) shows 60\% SIM success, pentagonal bipyramid (CSM = 0.88) achieves 67\%, and biaugmented trigonal prism (CSM = 3.37) also reaches 66\%, while capped trigonal prism (CSM = 1.72) has only 26\%. This non-monotonicity has a clear physical explanation: CSM measures deviation from an ideal polyhedron, but not all deviations are equivalent for SIM quality.

As an example, axial elongation of a square antiprism may increase the magnitude of the axial $B_2^0$ crystal field term as CSM increases. This behavior is consistent with the electrostatic model of Chilton and Soncini, where the local charge distribution at the ligand donor atoms is responsible for defining the anisotropic axis, not global polyhedral symmetry.\cite{Chilton2013}

\subsection{Generalization Across Families and Ions}

\begin{figure}[H]
\centering
\includegraphics[width=0.85\textwidth]{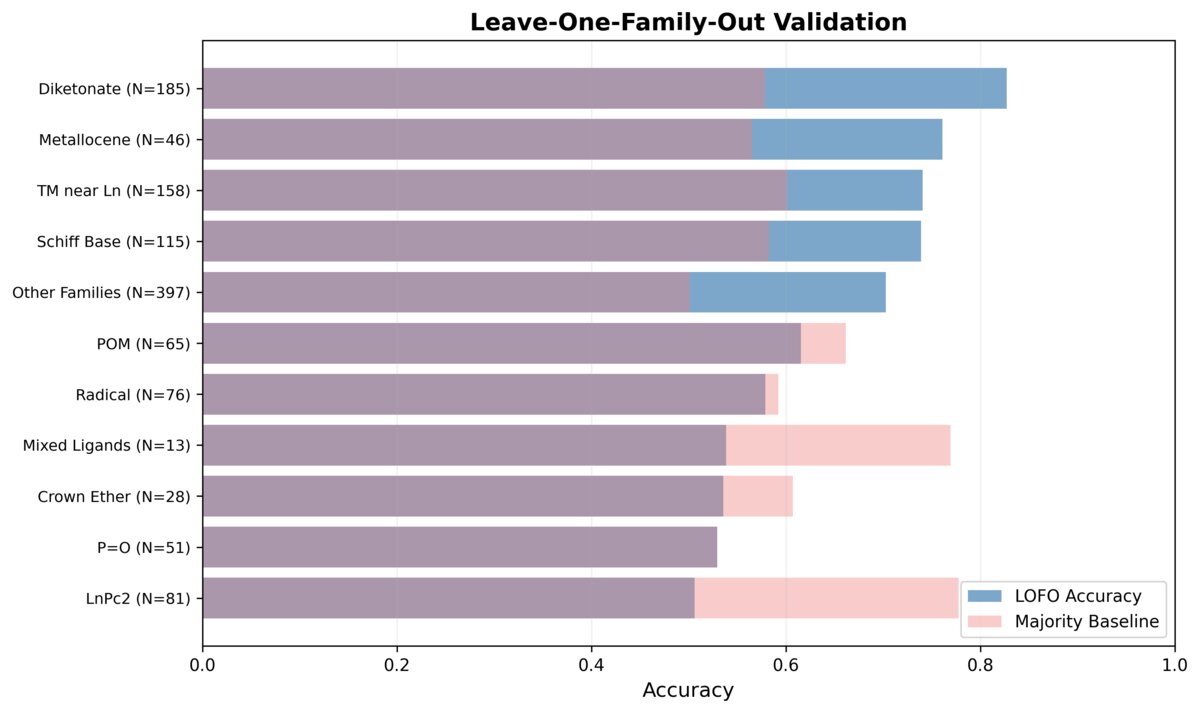}
\caption{Leave-one-family-out validation. Families are defined in Table~S1. The model generalizes to conventional families but fails for distinctive architectures (LnPc$_2$, P=O, Crown Ether).}
\label{fig:lofo}
\end{figure}

In Figure~\ref{fig:lofo}, we display leave-one-family-out accuracy. For structurally conventional families, the model generalizes well: diketonates (83\%), metallocenes (76\%), Schiff bases (74\%). For structurally distinctive families, the model fails: LnPc$_2$ (50.6\%), P=O (52.9\%), Crown Ether (53.6\%). The sandwich geometry of phthalocyanine complexes creates a qualitatively different crystal field environment that tabular descriptors do not adequately represent without the family label.

A model trained exclusively on Dy$^{3+}$ compounds ($N = 601$) achieves only 44\% mean accuracy when applied to other ions, compared to 76\% on held-out Dy$^{3+}$. Transfer learning thus largely fails, confirming that SIM design rules are ion-specific rather than universal.

\subsection{Prospective and Confidence-Stratified Validation}

\begin{figure}[H]
\centering
\includegraphics[width=0.85\textwidth]{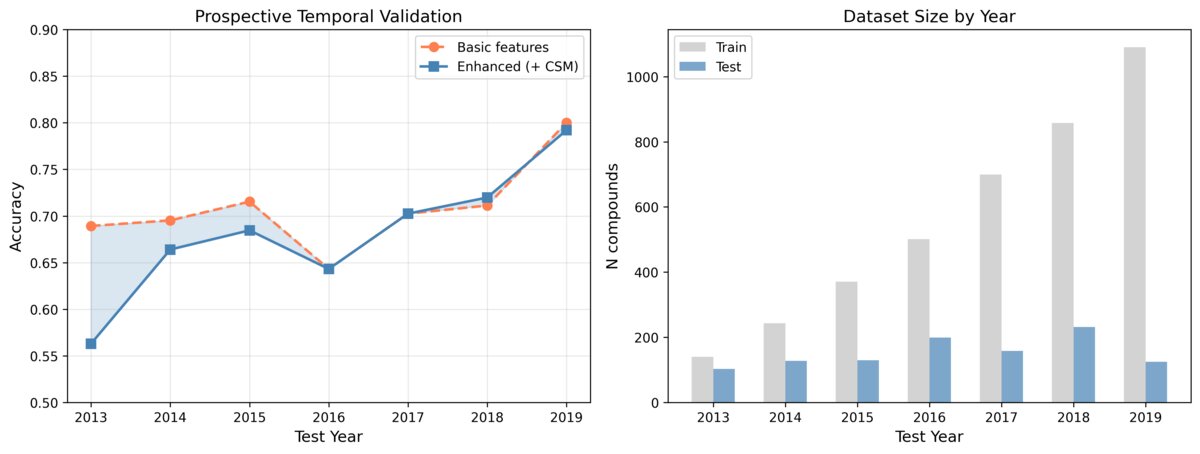}
\caption{Prospective temporal validation using actual publication years.}
\label{fig:temporal}
\end{figure}

Using actual publication years, we conducted a year-by-year prospective validation (Figure~\ref{fig:temporal}). We see an increase from 56\% for the 2013 test set (trained on 140 compounds) to 80\% for the 2019 test set (trained on 1,090 compounds). The mean temporal validation accuracy across the 2013--2019 period was 71\%, trailing standard cross-validation by only 5 percentage points. This gap highlights the expected distributional shift between earlier and later compounds.

\begin{figure}[H]
\centering
\includegraphics[width=0.7\textwidth]{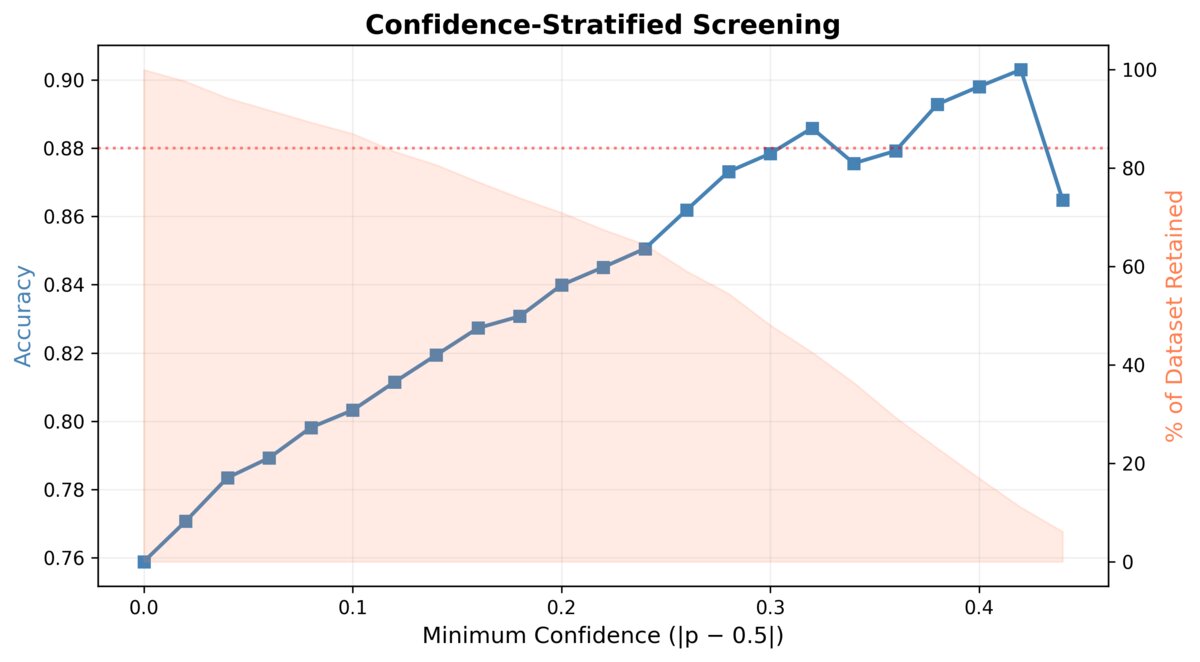}
\caption{Confidence-stratified accuracy. At $|p - 0.5| \geq 0.30$, accuracy reaches 88\% for 48\% of compounds.}
\label{fig:conf}
\end{figure}

For practical deployment, we evaluated accuracy as a function of prediction confidence (Figure~\ref{fig:conf}). At $|p - 0.5| \geq 0.30$, we retain 48\% of the database and reach accuracy levels of 88\%. This provides a directly actionable screening protocol: a synthetic chemist can set a confidence threshold matching their tolerance for false predictions.

\subsection{Misclassification Analysis and Post-Hoc Physics}

\begin{figure}[H]
\centering
\includegraphics[width=0.8\textwidth]{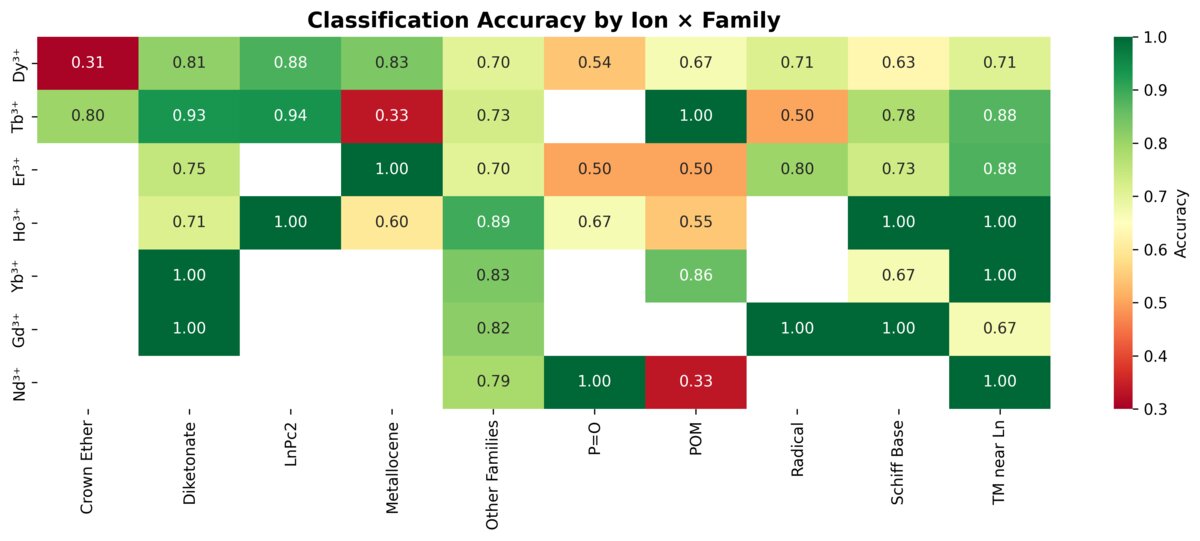}
\caption{Classification accuracy heatmap by ion $\times$ chemical family.}
\label{fig:heatmap}
\end{figure}

\begin{figure}[H]
\centering
\includegraphics[width=\textwidth]{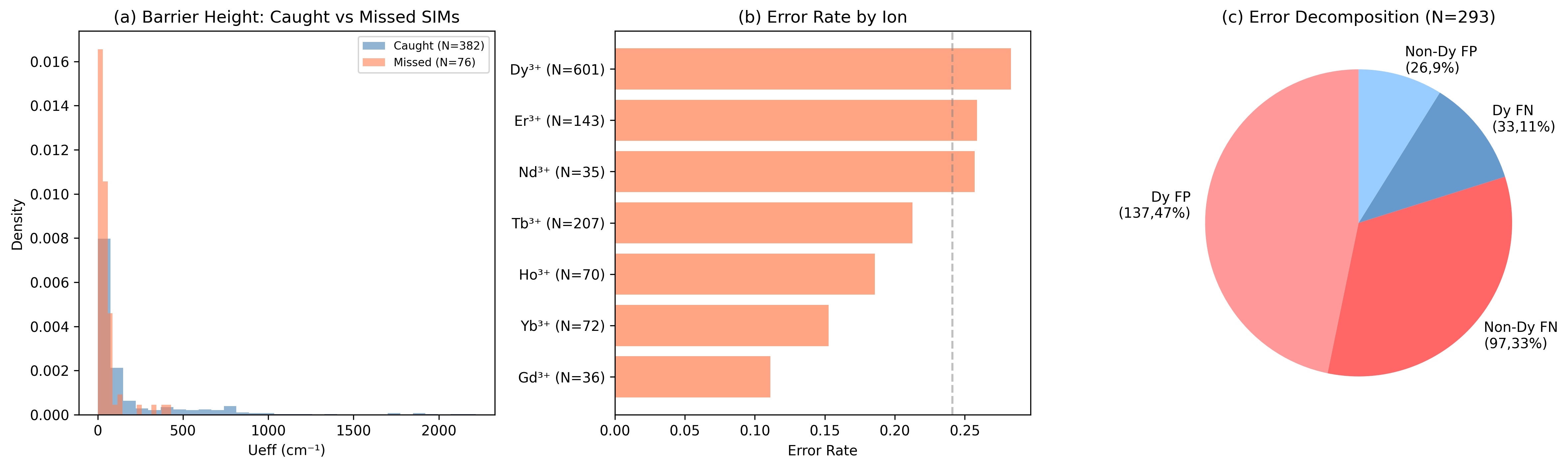}
\caption{Post-hoc analysis of misclassification. (a) $U_\text{eff}$ distributions for correctly classified and missed SIMs. (b) Error rates by ion. (c) Decomposition of all errors. Figure S15 includes additional panels demonstrating prediction confidence, CSM, and publication year.}
\label{fig:posthoc}
\end{figure}

Of the total 293 compounds misclassified, we note that there is a strongly asymmetrical error distribution (see Figures~\ref{fig:heatmap} and \ref{fig:posthoc}). False positives are overwhelmingly Dy$^{3+}$: 137/163 (84\%) are Dy compounds incorrectly predicted as SIM. False negatives cluster in non-Dy ions: the model misses 100\% of Yb$^{3+}$ and Gd$^{3+}$ SIMs.

Post-hoc analysis using relaxation parameters not available during training reveals that false negatives have significantly lower effective barriers: median $U_\text{eff} = 31.5$ cm$^{-1}$ versus 58.0 cm$^{-1}$ for correctly classified SIMs (Mann--Whitney $U$ test, $p < 0.0001$). This $1.8\times$ difference indicates that the model preferentially misses low-barrier SIMs---compounds exhibiting slow relaxation through Raman processes or phonon bottleneck effects rather than thermally activated Orbach relaxation. The error decomposition reveals that 80\% of all errors are attributable to the ion-identity bias: 47\% Dy$^{3+}$ false positives and 33\% non-Dy false negatives.

\subsection{Do 3D Crystal Structure Descriptors Break Through?}

We retrieved crystal structures from the Crystallography Open Database (COD) for 606 compounds and computed 49 three-dimensional descriptors: Coulomb matrix eigenvalues (15 features), coordination shell geometry (12 features), radial distribution function histogram (10 features), ligand--ligand distance statistics (3 features), and element counts (8 features).

\begin{figure}[H]
\centering
\includegraphics[width=\textwidth]{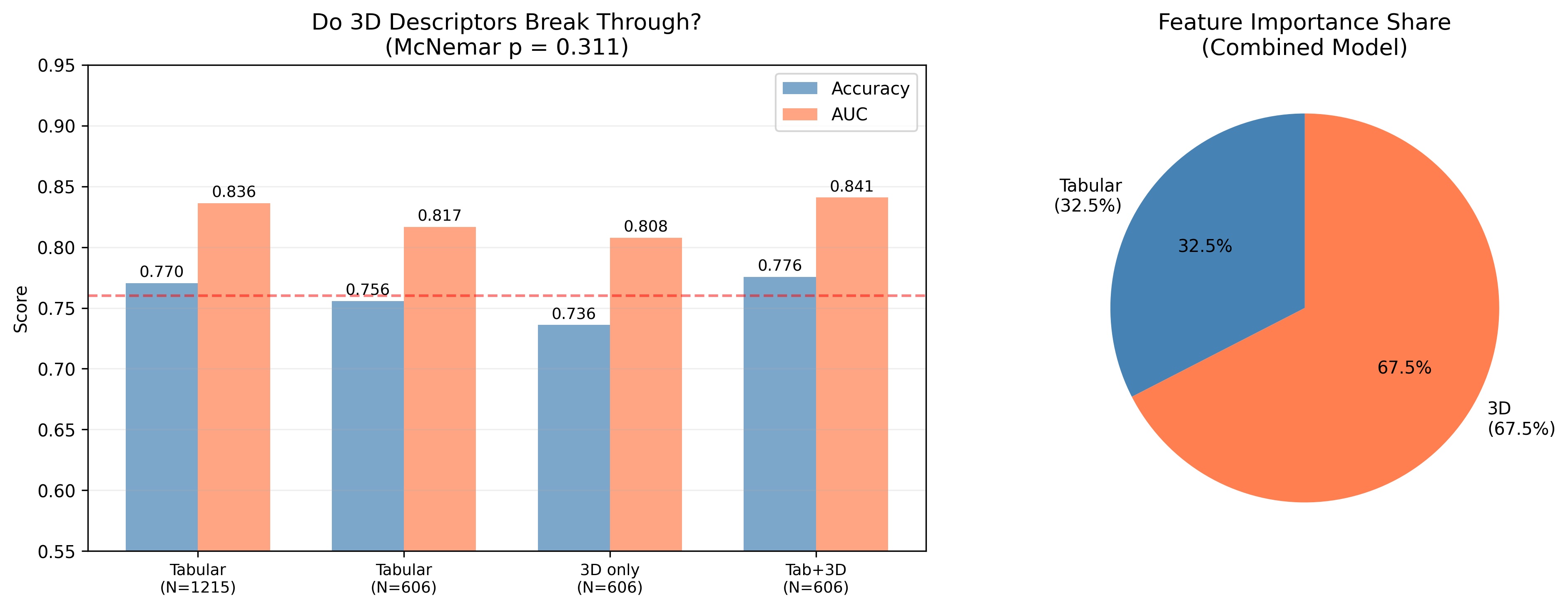}
\caption{3D crystal structure descriptors vs tabular baseline. Despite dominating feature importance (67.5\%), 3D coordinates do not significantly improve accuracy ($p = 0.31$).}
\label{fig:3d}
\end{figure}

In Table~\ref{tab:3d}, we summarize the results. Combining tabular and 3D features yields 77.6\% (AUC = 0.841), a $+2.0$ percentage-point improvement that is not statistically significant (McNemar $p = 0.31$). Critically, 17 of the top 20 features are 3D descriptors, yet this dominance in feature importance does not translate to improved classification. The 3D features encode the same predictive information in a different representation (Figure~\ref{fig:3d}).

Point-charge crystal-field descriptors derived from the same coordinates likewise provide no improvement: used alone, they yield 64\% classification accuracy, while adding them to the tabular features reduces accuracy from 75.7\% to 71.8\% (Supporting Information, Section S19). This convergence across three levels of geometric description---tabular, CSM, and 3D coordinates---provides strong evidence that the bottleneck is not geometric precision but the absence of electronic structure information.

\begin{table}[H]
\centering
\caption{3D crystal structure descriptors vs tabular baseline. McNemar $p$(Tabular vs Tabular+3D) = 0.31. The full-dataset baseline was recomputed within the 3D pipeline with the same hyperparameters and seed; the difference from Table~\ref{tab:baselines} (0.759) lies within the fold-to-fold variability of the cross-validation (0.75--0.77 across random seeds).}
\label{tab:3d}
\begin{tabular}{lcccc}
\toprule
Model & $N_\text{feat}$ & Accuracy & AUC & $N$ \\
\midrule
Tabular (full dataset) & 30 & 0.770 & 0.836 & 1215 \\
Tabular (COD subset) & 30 & 0.756 & 0.817 & 606 \\
3D descriptors only & 49 & 0.736 & 0.808 & 606 \\
Tabular + 3D & 79 & 0.776 & 0.841 & 606 \\
\bottomrule
\end{tabular}
\end{table}

\subsection{Ab Initio and Structural Validation of Error Mechanisms}
\label{sec:abinitio}

To understand what information is missing from the descriptor space, we reviewed several representatives of the two error classes identified in Section 3.6. Specifically, we evaluated representative examples of false-negative classifications, i.e., SIMs that were not identified by the model, and assessed how well the magnetic properties of high-confidence false positives can be explained by means of interactions beyond the single ion.

\subsubsection{False negatives: Ab initio anisotropy of missed SIMs}

From the false-negative pool, we selected several mononuclear Yb$^{3+}$ complexes with ordered crystal structures listed in the Crystallography Open Database (COD). Yb$^{3+}$ is the ion class for which the model failed to identify every experimentally defined SIM (see Section 3.6). For each complex, we identified the molecular subunit in question from the crystal structure, assigned charges based upon the requirement of overall neutrality of the crystallographically defined formula unit, and verified our assignments via parity of electron counts. Each complex was then subjected to state-averaged CASSCF(13,7) calculations over all seven doublet states associated with the $^2$F term using the exact two-component (X2C) relativistic Hamiltonian, atomic mean-field spin--orbit integrals, and Cholesky-decomposed two-electron integrals as implemented in OpenMolcas. Spin--orbit coupling was included via RASSI-SO, and the $g$ tensors of the resulting Kramers doublets were computed via SINGLE\_ANISO.\cite{OpenMolcas2019,Chibotaru2012} Throughout, a tiered ANO-RCC basis was used: VTZP on Yb, VDZP on donor atoms immediately coordinating with Yb, and VDZ on all remaining carbon and hydrogen atoms.

An N$_4$O-donor Yb chelate containing three coordinated nitrate groups and possessing a coordination number of 11 (COD 4349807) represents a highly illustrative case. The model assigned $p$(SIM) = 0.115 to this compound, which makes it one of the most confident errors made by the model. However, this compound has been shown experimentally to display slow magnetic relaxation ($T_{B3} > 2$~K). The calculated electronic structure describes why this compound was not recognized by the structural model and why it is a relatively poor SIM.

The $^2$F$_{7/2}$ ground multiplet splits into four Kramers doublets at 0, 70, 176, and 426 cm$^{-1}$. The ground Kramers doublet is very strongly axial, with $g_z = 7.18$; in comparison, the Ising limit for a pure $|m_J = \pm 7/2\rangle$ doublet would correspond to $g_z = 8$. Thus, this compound displays the easy-axis anisotropy associated with a SIM. On the other hand, the transverse components of this doublet are not negligible, with $g_x = 0.65$ and $g_y = 1.06$, and the first excited doublet lies only 70 cm$^{-1}$ above the ground state. As such, these non-negligible transverse contributions indicate that there is considerable potential for efficient ground-state quantum tunnelling.

This combination of significant axial anisotropy and substantial transverse anisotropy, along with a close-lying excited doublet, is indicative of a weak, field-induced relaxer rather than a strong, thermally activated Orbach magnet. Therefore, COD 4349807 illustrates a microscopically observable realization of the statistically significant trend identified in Section 3.6, in which missed SIMs had barriers 1.8 times lower than those of correctly classified SIMs. Neither the tabular nor the geometric descriptors capture the transverse anisotropy of the ground Kramers doublet, because this anisotropy arises from the crystal-field wavefunction rather than simply being a function of the coordination geometry.

A second Yb$^{3+}$-based complex from the same synthetic study was examined to provide further corroboration. The cationic N-donor cage (COD 4349809) contains two coordinated triflate groups and was assigned a model confidence of $p$(SIM) = 0.090 despite displaying slow magnetic relaxation. Its calculated anisotropy is similar to that displayed by COD 4349807. Four Kramers doublets are found at 0, 103, 253, and 333 cm$^{-1}$, and the $g$ tensor for the ground doublet is given by $g_x = 0.70$, $g_y = 1.07$, and $g_z = 6.88$.

Therefore, two independently misclassified Yb$^{3+}$ compounds that differ significantly in their coordination geometries display virtually identical electronic-structure signatures: a moderately axial ground doublet accompanied by transverse $g$ components that are approximately unity. The recurrence of this pattern indicates that tunnelling-prone, field-induced relaxation is likely to be a common feature among the model's most confident Yb$^{3+}$ false negatives.

As a positive control, we investigated COD 7107749, an anionic N$_8$ double-decker Dy$^{3+}$ complex that was correctly classified by the model with $p$(SIM) = 0.878. To evaluate this compound computationally, we performed CASSCF(9,7) calculations, state-averaging over the 21 sextet states, and subsequently applied RASSI-SO to treat the resulting $^6$H$_{15/2}$ manifold. The CASSCF/RASSI-SO/SINGLE\_ANISO protocol was again employed using the X2C Hamiltonian and ANO-RCC basis sets.

There exists a considerable difference between the electronic-structure characteristics of the correctly classified control and the Yb$^{3+}$ false negatives. The eight Kramers doublets of the $^6$H$_{15/2}$ multiplet span 776 cm$^{-1}$ and occur at 0, 83, 91, 211, 335, 349, 726, and 776 cm$^{-1}$. The ground doublet has a large axial component, $g_z = 16.9$, or roughly 85\% of that expected for a pure $|m_J = \pm 15/2\rangle$ state, while its transverse components are much smaller, $g_x = 0.71$ and $g_y = 0.75$. Clearly, then, the same electronic-structure protocol can distinguish between the correctly classified control and the Yb$^{3+}$ false negatives. The shallow level structure and large transverse anisotropy of the latter thus appear to be intrinsic to these compounds rather than artifacts of the computational treatment.

Additional tests of robustness were conducted on COD 4349807 using four-state multistate CASPT2 calculations employing IPEA = 0 and an imaginary shift of 0.1. The axial component was altered by only approximately 4\% ($g_z = 7.18$ versus $g_z = 7.46$), and clearly, the transverse components remained finite. Thus, the marginal nature of the ground doublet is not an artifact of the CASSCF treatment. Additionally, these calculations were reproduced to all reported digits on two different computers employing versions v26.02 and v26.06 of OpenMolcas, as documented in the Supporting Information.

\subsubsection{High-confidence false positives: Structures beyond the single-ion paradigm}

In addition to providing a mechanistic explanation for the two error types discussed in Section 3.6, these calculations and structural analyses provide insight into whether the false positives share any structural or magnetic features that might preclude them from fitting into a purely local, single-ion picture. Multinuclear assemblies, exchange-coupled radicals, and intermolecular spin networks can generate relaxation mechanisms that are not accounted for in descriptors that rely exclusively on the coordination environment about an individual lanthanide ion.

We investigated whether crystal structures could be obtained for all false positives with high-confidence assignments. Each structure was classified as either mononuclear or multinuclear depending on whether or not its connected molecular unit contained more than one 4f center. Heterometallic systems and radical-bearing systems were individually identified as such; we detected radical units through identification of N--O bond connections indicative of nitroxide functionality.

These studies indicate that the features associated with many of the top false positives fit well with an interpretation involving non-local relaxation mechanisms. All three of the highest-confidence false positives contain aspects that extend beyond an isolated single-ion representation: a trinuclear Dy$_3$ cluster with $p$(SIM) = 0.94, an extended Dy--Y dinuclear system with $p$(SIM) = 0.92, and a Dy--nitronyl-nitroxide radical complex with $p$(SIM) = 0.88. The last of these compounds belongs to a previously established family of exchange-coupled 4f--radical systems wherein direct Dy--radical interactions contribute to the relaxation dynamics, and hence isolated-ion descriptions of the crystal field are incomplete.

Four of the eight high-confidence false positives for which structures were identifiable are multinuclear, metal-bridged, or radical-bearing. The remaining four are genuinely mononuclear and include an organometallic sandwich complex belonging to one of the best-known structural families of SIMs. This demonstrates another type of constraint associated with the descriptor space: local geometry alone is insufficient for predicting relaxation; i.e., local geometry may favorably describe a SIM-like coordination environment yet lack sufficient detail to predict actual slow relaxation.

The nearest-neighbor separations for the false positives tend to follow this same logic: the false positives have a slightly shorter median Dy$\cdots$Dy separation than the true positives, i.e., 7.9 versus 9.7~\AA, although no statistical significance can be attached to this observation at the available pool sizes (one-sided Mann--Whitney $p = 0.14$). Furthermore, crystal structures could not be obtained for seven of the fifteen high-confidence false positives; therefore, only the structurally classifiable subpopulation can be considered herein.

Together, these calculations and structural investigations provide evidence supporting an understanding of how the false negatives and false positives arise in terms of failure modes: the false negatives appear to result primarily from missing wavefunction-level crystal-field anisotropies, whereas the false positives appear to arise from missing structural interaction topologies beyond the individual metal center, specifically multinuclear, metal-bridged, or radical-bearing assemblies, and from missing information regarding potentially relevant dynamical factors contributing toward slow relaxation.

Therefore, these results support recommendations for the future development of improved models: the inclusion of wavefunction-level crystal-field anisotropic parameters; the inclusion of structural features describing interaction topologies beyond the individual metal center, such as multinuclear, metal-bridged, or radical-bearing assemblies; and the incorporation of relevant dynamical factors, such as potential contributions from intermolecular dipole--dipole interactions. Additionally, some or all of this new information could be incorporated as ab initio features or as inexpensive structural filtering protocols.

\begin{table}[H]
\centering
\caption{Ab initio validation set. $p$(SIM) denotes the model's out-of-fold prediction; experimental labels are from SIMDAVIS. Ground-Kramers-doublet $g$ tensors were obtained with CASSCF/RASSI-SO/SINGLE\_ANISO (X2C, ANO-RCC).}
\label{tab:abinitio}
\begin{tabular}{lccccc}
\toprule
COD & Ion & $p$(SIM) & Expt. & ($g_x$, $g_y$, $g_z$) & KD1$\to$KD2 (cm$^{-1}$) \\
\midrule
4349807 & Yb$^{3+}$ & 0.115 & SIM ($T_{B3} > 2$ K) & (0.65, 1.06, 7.18) & 70 \\
4349809 & Yb$^{3+}$ & 0.09 & SIM ($T_{B3} > 2$ K) & (0.70, 1.07, 6.88) & 103 \\
7107749 & Dy$^{3+}$ & 0.878 & SIM (control) & (0.71, 0.75, 16.88) & 83 \\
\bottomrule
\end{tabular}
\end{table}

\section{Conclusion}

Based on a systematic analysis of the SIMDAVIS 1.2.1 database, this study evaluates how effectively coordination-level structural descriptors predict single-ion magnet (SIM) behavior in lanthanide complexes. Key findings indicate that simple ion identity (e.g., Dy$^{3+}$) dominates prediction with 71\% accuracy, while complex gradient boosting using 30 features only increases this to 76\%, and adding 18 Continuous Symmetry Measure (CSM) features offers no significant improvement, demonstrating that more structural data does not necessarily aid prediction. CSM is largely redundant because it measures the magnitude of symmetry breaking rather than its alignment with the crystal field axis.

Generalization continues to be unbalanced. Conventional families such as diketonates have high accuracy levels (e.g., diketonates had an accuracy level of 83\%), while distinctive architectural families have low accuracy levels (e.g., LnPc$_2$ had an accuracy level of 51\%). We found that ion-identity bias accounted for $\sim$80\% of all misclassifications, and we demonstrated that missed SIMs had median $U_\text{eff}$ values that were on average $1.8\times$ less than correctly classified SIMs ($p < 0.0001$), which is consistent with weak, field-induced relaxation that static structural descriptors cannot accurately describe.

When trained exclusively on Dy$^{3+}$, we observed that the classifier reached only 44\% accuracy on other ions, demonstrating that transferable SIM design rules must consider explicit ion-specific electronic structure. The proof-of-concept multireference ab initio calculations and structural analysis of representative errors (Section~\ref{sec:abinitio}) support this interpretation: the false-negative cases that we examined included large transverse ground-doublet components, while many leading false-positive candidates include multinuclear or radical-bearing systems outside of the single-ion paradigm.

While we acknowledge these limitations, we conclude that confidence-stratified screening will prove to be practically useful: at $|p - 0.5| \geq 0.30$, we achieved 88\% accuracy while retaining 48\% of the compounds. Conversely, adding a further 49 three-dimensional descriptors describing the 606 COD structures improved accuracy by only 2.0 percentage points, indicating that the major barrier lies electronically rather than geometrically. Therefore, we anticipate improvements to this limit will require integration of structural databases with computational chemistry methods. Based upon this perspective, we propose that inclusion of CASSCF-derived crystal field splittings and ground-doublet $g$ tensors would be the most direct follow-on features, supplemented by inexpensive filters for nuclearity and radical character.

\section*{Author contributions}
F.Z.: conceptualization, methodology, software, investigation (ab initio calculations), formal analysis, writing (original draft, review and editing). R.P.: software, data curation, formal analysis, visualization, writing (review and editing). V.-A.G.: conceptualization, methodology, validation, formal analysis, writing (review and editing). D.P.: conceptualization, methodology, investigation, supervision, funding acquisition, writing (review and editing). All authors have read and approved the final manuscript.

\section*{Conflicts of interest}
There are no conflicts to declare.

\section*{Data availability}
This study was carried out using the publicly available SIMDAVIS 1.2.1 database (release r2025\_02\_18) at \url{https://rosaleny.shinyapps.io/simdavis_dashboard/}.\cite{Duan2022,Canon2025} The scripts that build the feature sets, train and evaluate the classifiers, and produce every figure and table, together with the crystal structures used, the outputs of all analyses, and the OpenMolcas input and output files, are available on Zenodo at \url{https://doi.org/10.5281/zenodo.22676548}. Additional analyses have been included as part of the Supporting Information, which follows the main text in this preprint.

\begin{acknowledgement}
Part of this work was supported by the U.S. Department of Energy, Office of Science, under Contract No.\ DE-AC02-07CH11358 at Ames National Laboratory. The machine-learning work was supported as part of the Center for Energy Efficient Magnonics, an Energy Frontier Research Center funded by the U.S.\ Department of Energy, Office of Science, Basic Energy Sciences, under Award No.\ DE-AC02-76SF00515. F.Z. and V.-A.G. acknowledge support from the Laboratory Directed Research and Development (LDRD) program of Oak Ridge National Laboratory (LOIS ID 12735). This research used resources of the Oak Ridge Leadership Computing Facility (allocation CHM238), a DOE Office of Science User Facility operated under Contract No.\ DE-AC05-00OR22725, and of the National Energy Research Scientific Computing Center (NERSC), a DOE Office of Science User Facility operated under Contract No.\ DE-AC02-05CH11231, under award m4621 (ERCAP0036406). We acknowledge Ca\~{n}\'{o}n-Mancisidor, Gonz\'{a}lez Ponce, Rosaleny, and Gaita-Ari\~{n}o\cite{Canon2025} and the SIMDAVIS team\cite{Duan2022} for making the SIMDAVIS database publicly available.
\end{acknowledgement}

\begin{suppinfo}
Complete feature lists and hyperparameter settings; physics-based rule baselines; classifier comparisons; three-class analysis; cross-ion transfer learning; Orbach residuals; a summary figure; and additional SHAP analyses. The Supporting Information is appended after the reference list in this preprint.
\end{suppinfo}

\bibliography{references}

\clearpage


\section*{Supplementary material}

# Supporting Information: From Heuristics to Machine Learning: The Performance Ceiling for Single-Ion Magnets and Its Electronic Origin

Federico Zahariev,*,† Regina Pereyra,† Vassiliki-Alexandra Glezakou,*,‡ and Durga Paudyal*,¶

†*Ames National Laboratory and Department of Chemistry, Iowa State University, Ames, Iowa 50011, USA*

‡*Chemical Sciences Division, Oak Ridge National Laboratory, Oak Ridge, Tennessee 37830, USA*

¶*Department of Physics and Astronomy, University of Iowa, Iowa City, Iowa 52242, USA*

E-mail: fzahari@iastate.edu; glezakouva@ornl.gov; durga.quantum@gmail.com

## 1 Chemical Families

Table S1 defines the 11 chemical families of SIMDAVIS 1.2.1 used in Figure 4 of the main text. A family is assigned from the ligands coordinated to the lanthanide ion.

Table S1: Chemical families in the classification set ($N = 1215$).

| Family | $N$ | Definition |
|---|---|---|
| Diketonate | 185 | At least two $\beta$-diketonate ligands |
| TM near Ln | 158 | A transition metal ion near the lanthanide |
| Schiff base | 115 | Schiff base ligands |
| $LnPc_2$ | 81 | Bis(phthalocyaninato) double-decker sandwich |
| Radical | 76 | Radical-containing ligands |
| POM | 65 | At least one polyoxometalate ligand |
| P=O | 51 | At least one oxygen donor from a P=O (phosphonate-type) ligand |
| Metallocene | 46 | Cyclopentadienyl-type sandwich |
| Crown ether | 28 | A crown ether in the coordination sphere |
| Mixed ligands | 13 | Several ligand types, none dominant |
| Other families | 397 | Outside the preceding categories |

# 2 Complete Feature Set

The complete list of characteristics (30) consists of: numerical (3): coordination number (CN), number of ligands ($n_{\mathrm{lig}}$), concentration (%); anisotropy (2): is_oblate, is_prolate (binary); electronic (1): is_kramers (binary); elements of coordination (3): has_oxygen, has_nitrogen, has_carbon (binary); chemical family (11): one-hot encoded; ion identity (10): one-hot encoded for Ce, Dy, Er, Gd, Ho, Nd, Pr, Tb, Tm, Yb.

In addition to the above characteristic list, the extended list of characteristics (48) also includes: CSM (1), CSM_missing (1), axial_distortion (1), axial_dist_missing (1), slope_CSM (1), closest polyhedra (10, one-hot), poly_missing (1), and interaction terms (2): CSM×oblate, axial_distortion×Kramers.

Table S2 gives short definitions of the feature labels as they appear in the SHAP plots (Figure 2 of the main text and Figures S6–S10).

Table S2: Feature labels used in the SHAP plots.

| Label | Definition |
|---|---|
| ion_X | 1 if the lanthanide ion is X (Dy, Er, ...), else 0 |
| fam_X | 1 if the compound belongs to family X (Table S1), else 0 |
| poly_X | 1 if the closest ideal polyhedron from SHAPE is X, else 0 |
| CN | coordination number |
| n_lig | number of ligands |
| has_oxygen (_nitrogen, _carbon) | 1 if at least one donor atom is O, N, or C, else 0 |
| is_oblate, is_prolate | 1 if the $4f$ electron density of the ion is oblate (prolate), else 0 |
| is_kramers | 1 if the ion has an odd number of $4f$ electrons, else 0 |
| CSM | continuous symmetry measure of the coordination polyhedron with respect to the closest ideal polyhedron (SHAPE); 0 for an ideal polyhedron |
| axial_dist | axial distortion of the coordination polyhedron, as tabulated in SIMDAVIS |
| slope_CSM | CSM slope, as tabulated in SIMDAVIS |
| CSM_x_oblate | CSM × is_oblate |
| axdist_x_kramers | axial_dist × is_kramers |
| _missing | 1 if the corresponding SHAPE quantity is unavailable and was imputed, else 0 |

# 3 Hyperparameter Settings

Gradient boosting: $n_{\mathrm{estimators}} = 200$, max_depth = 4, learning_rate = 0.05, min_samples_leaf = 5, random_state = 42. Random forest: $n_{\mathrm{estimators}} = 200$, max_depth = 8, min_samples_leaf = 3. SVM: RBF kernel, $C = 1.0$, $\gamma$ = 'scale'. MLP: hidden_layer_sizes = $(64, 32)$, max_iter = 500, early_stopping = True. All features were normalized to the range $[0, 1]$ using min–max scaling.

# 4 Classifier Comparison

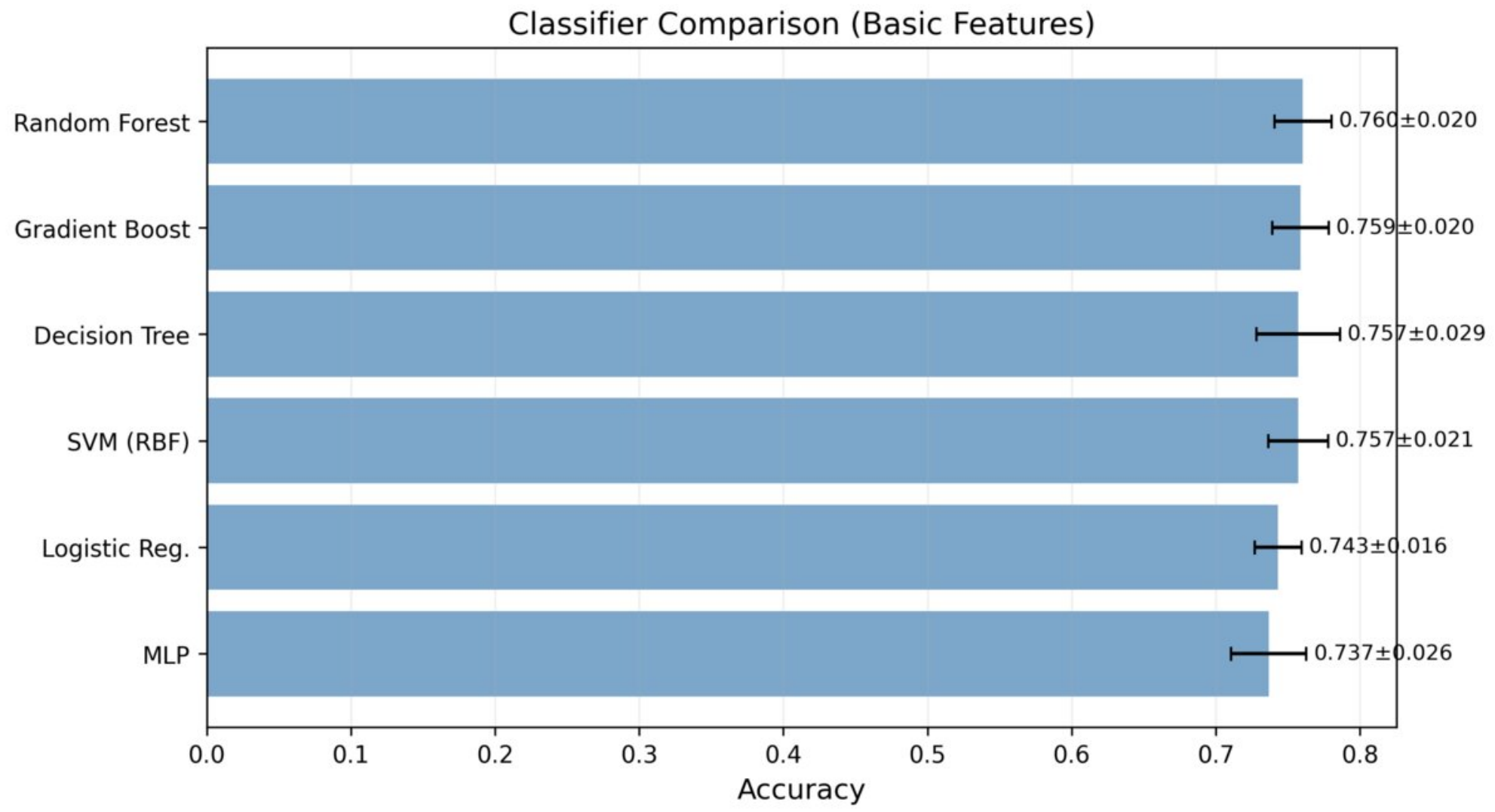


Figure S1: Classifier comparison for the basic features (5-fold CV, $N = 1215$). All classifiers converge close to 73–76% accuracy.

## 5 Three-Class Analysis

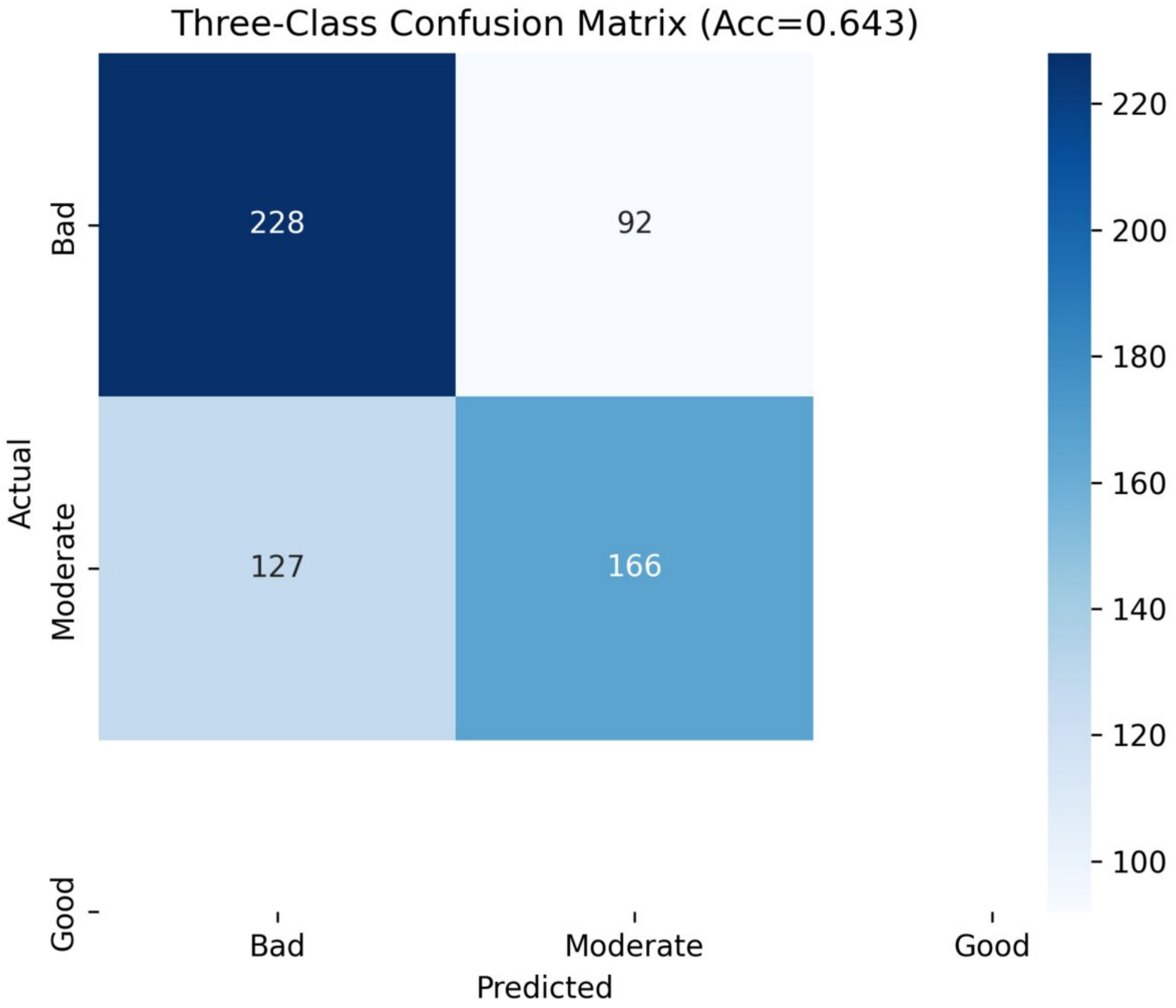


Figure S2: Confusion matrix for three classes (no $\chi''$; frequency-dependent $\chi''$; $T_{B3} > 2$ K).

# 6 Cross-Ion Transfer Learning

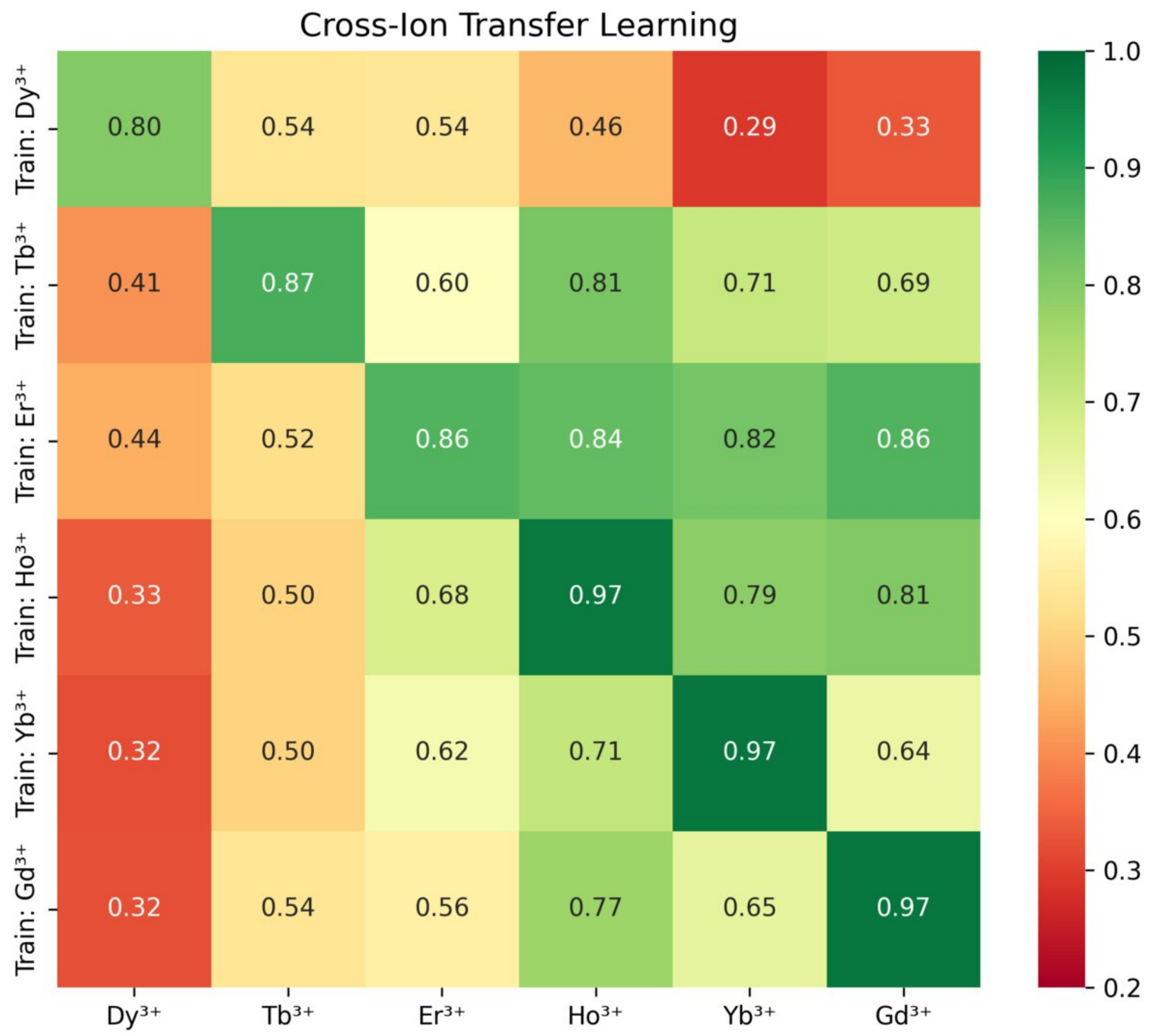


Figure S3: Accuracy matrix for cross-ion transfer learning. Models trained on one ion and tested on other ions. Dy→others: 44% average.

# 7 Orbach Residuals

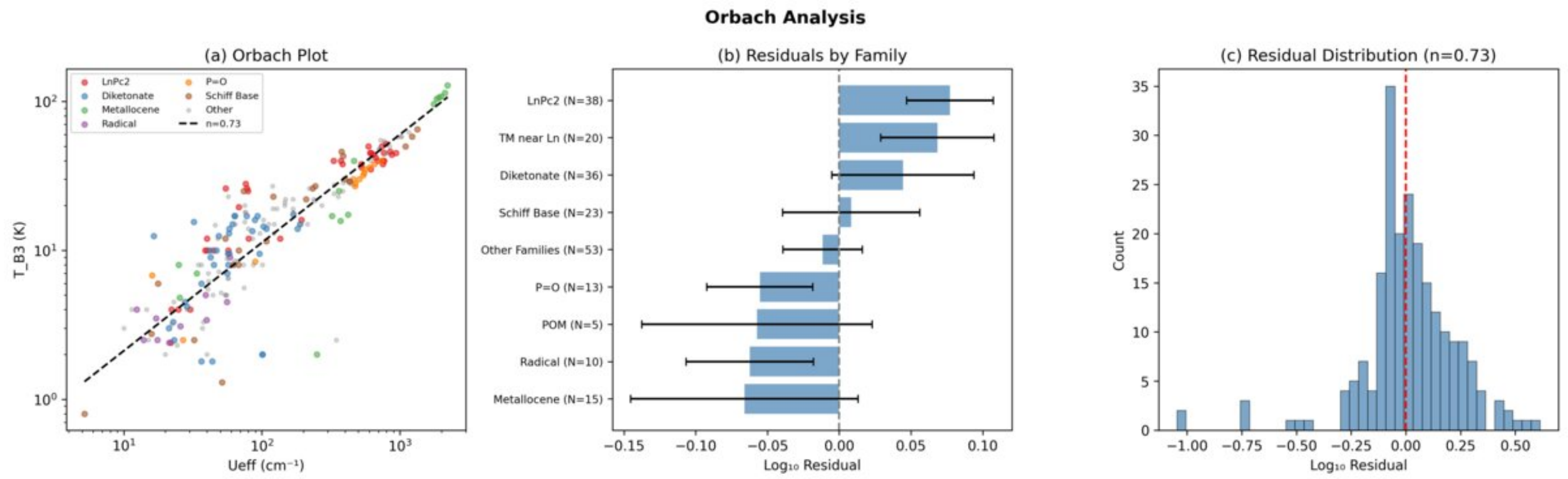


Figure S4: Orbach fit residuals ($T_{B3}$ versus $U_{\mathrm{eff}}$) by family and ion.

# 8 Grand Summary

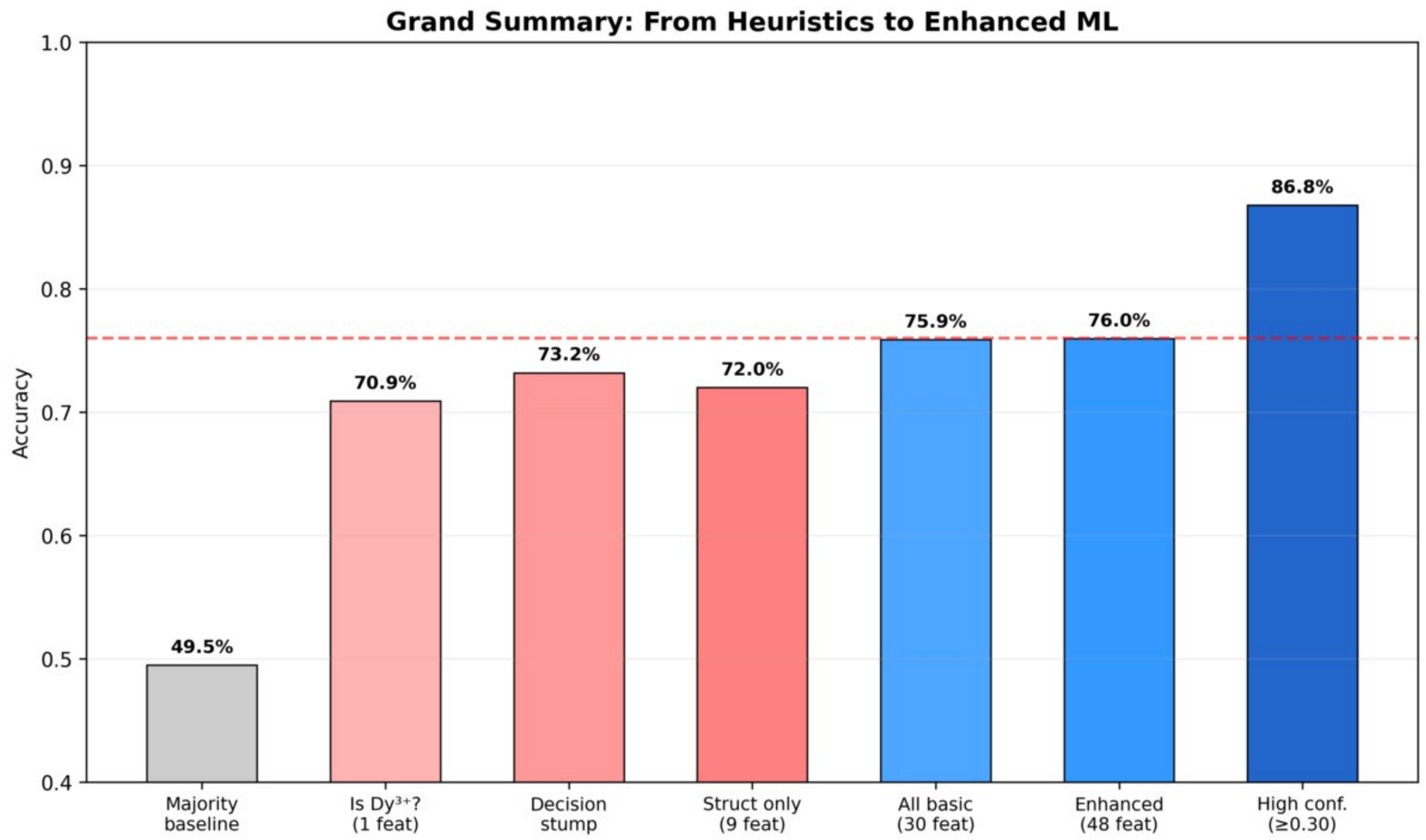


Figure S5: Overall summary of all validation protocols, showing that accuracy converges to approximately 76%.

# 9 SHAP Analysis — Basic Model

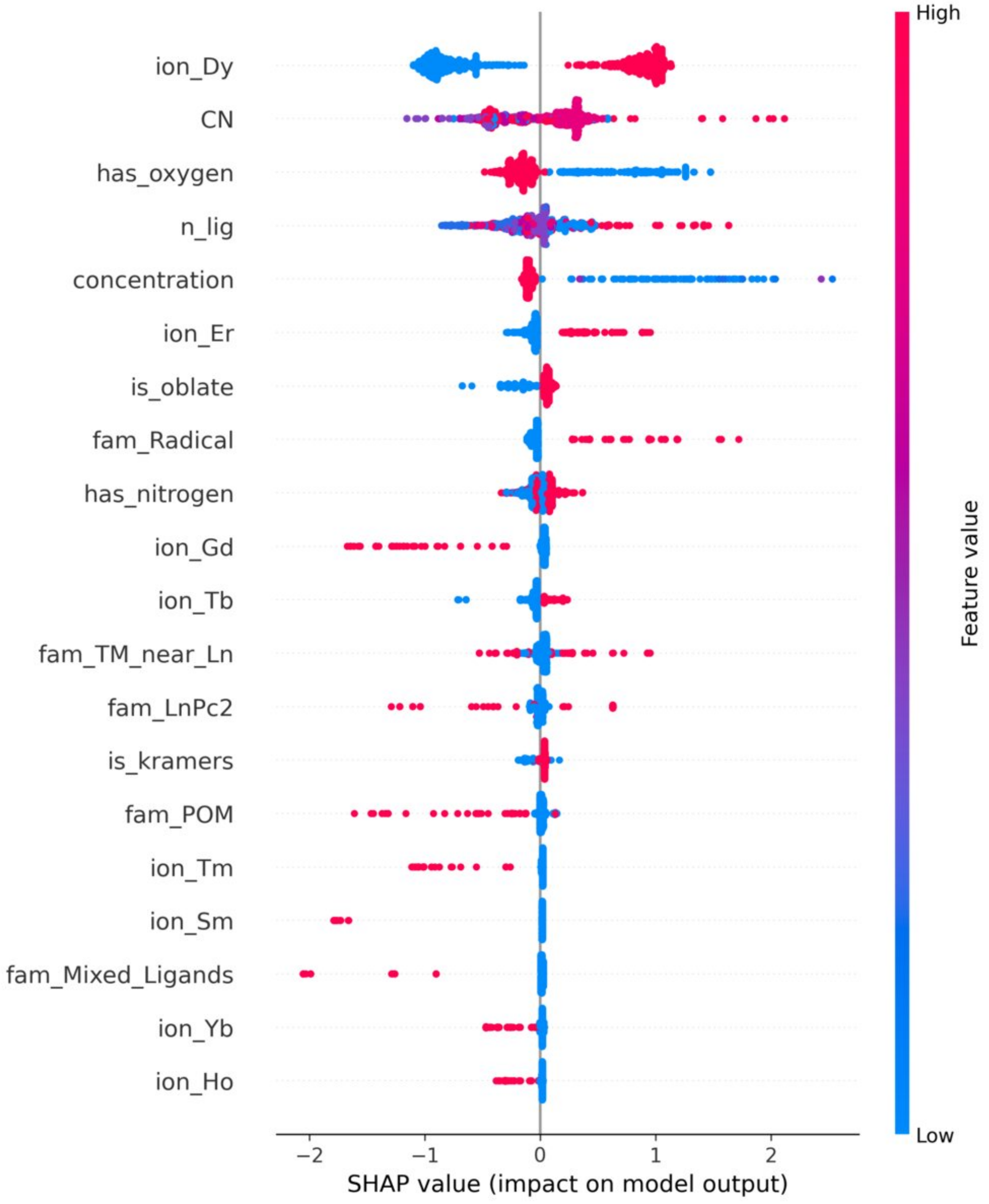


Figure S6: SHAP beeswarm for the basic model (30 features, $N = 1215$).

# 10 SHAP Waterfall Examples

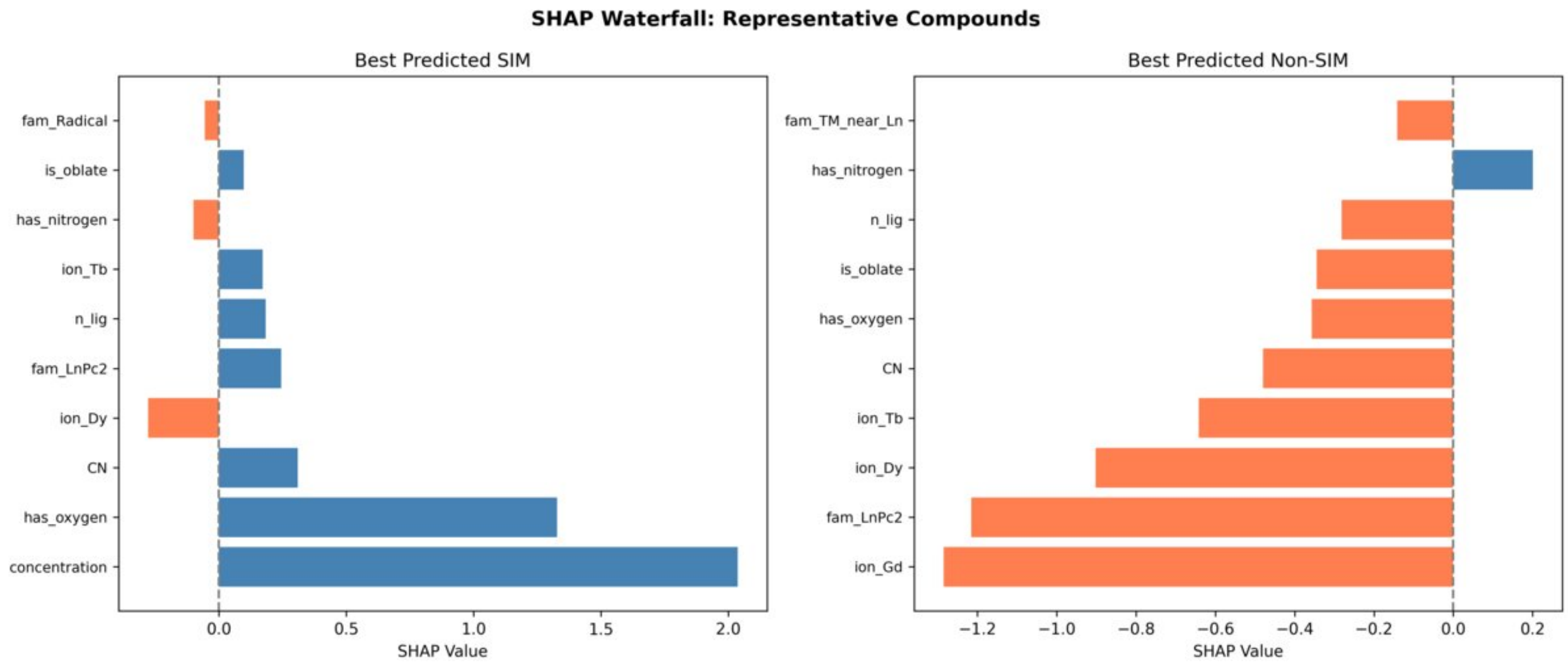


Figure S7: SHAP waterfall plots for representative compounds displaying the individual contribution of each feature to the prediction.

# 11 SHAP Dependence: CN and Oxygen

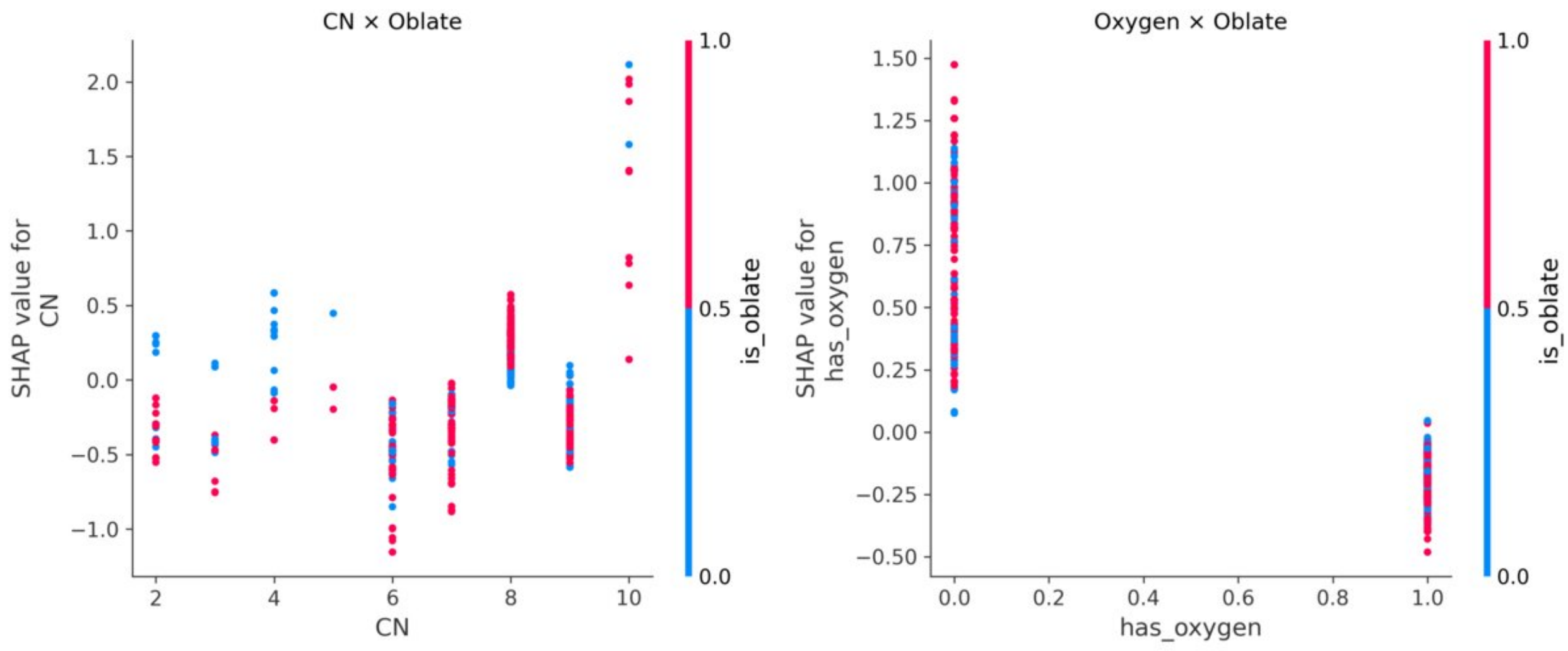


Figure S8: SHAP dependence plots for coordination number and oxygen presence.

# 12 Average SHAP by Family

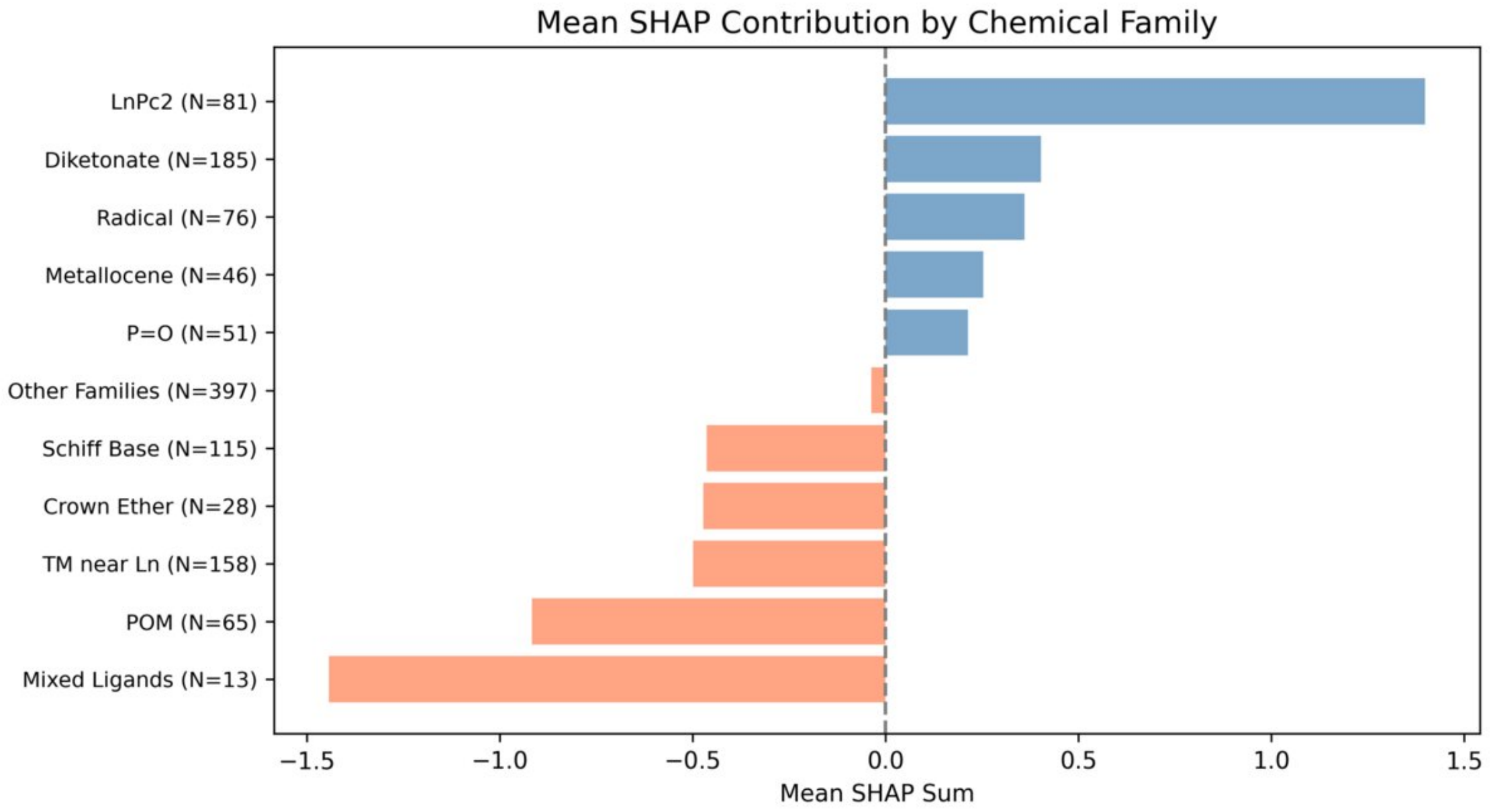


Figure S9: Mean absolute value of SHAP per family to identify family-specific patterns of feature importance.

# 13 SHAP Interaction: CN × Oblate

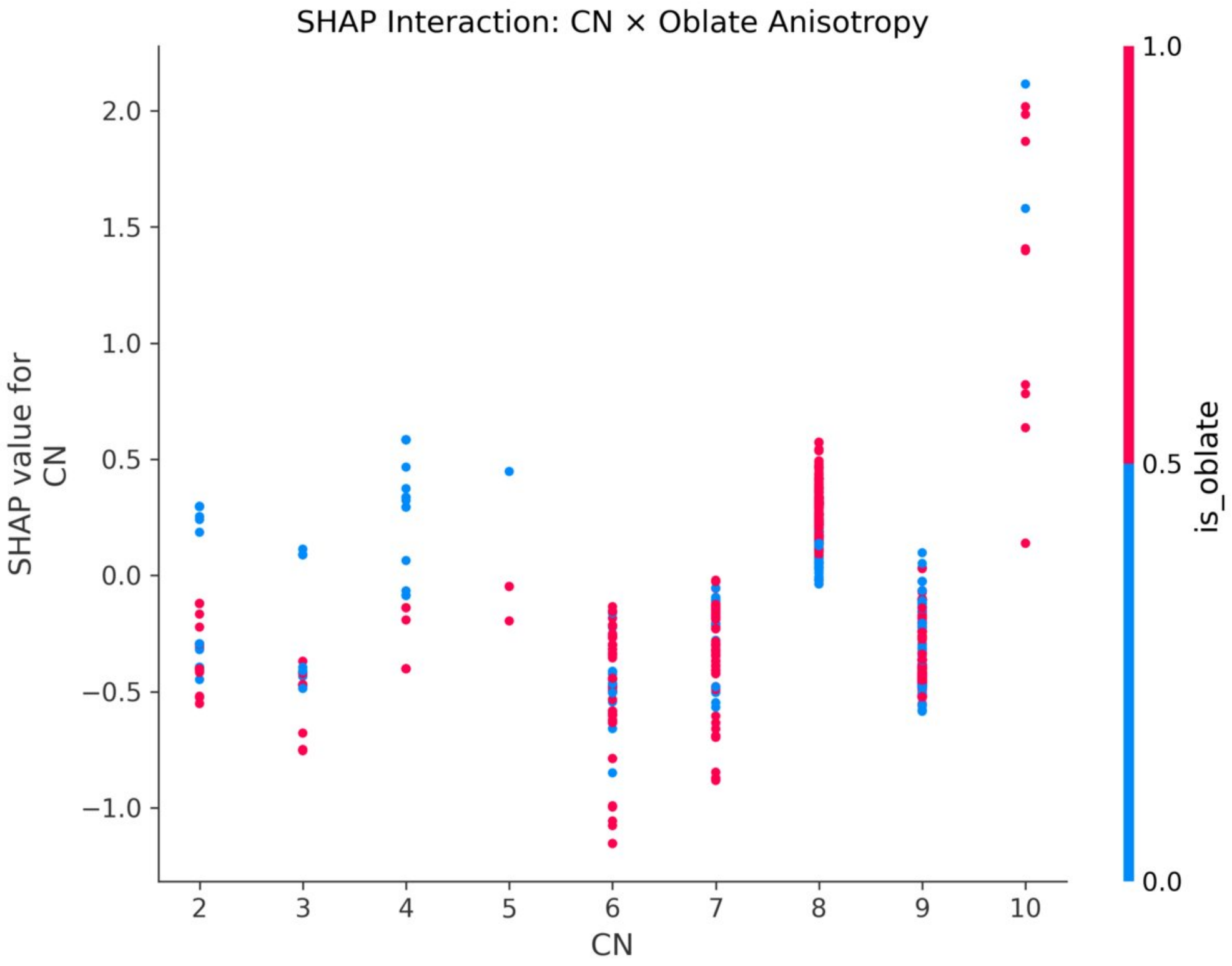


Figure S10: SHAP interaction values for CN × is_oblate.

# 14 Model Comparison Bar Chart

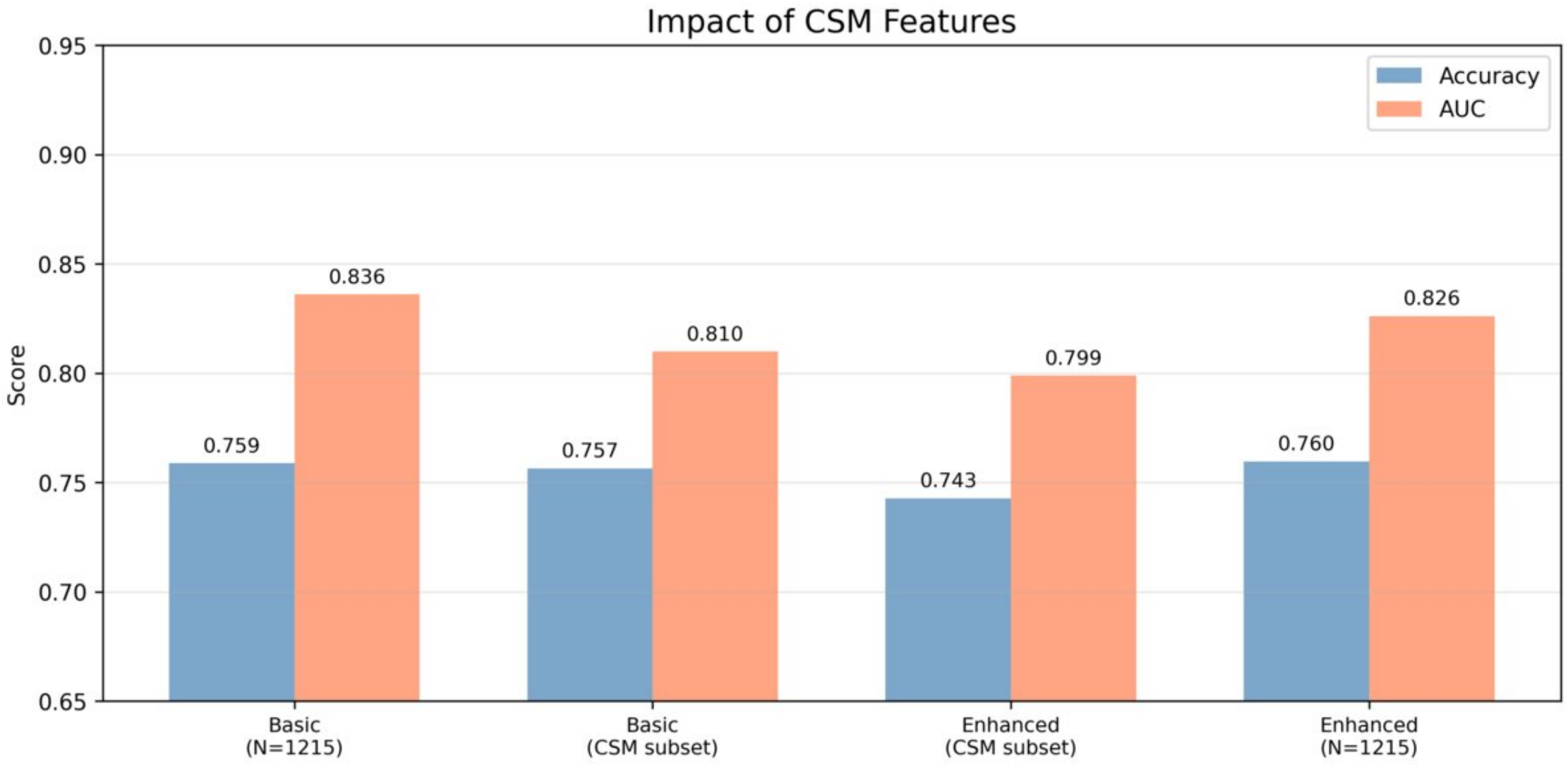


Figure S11: Detailed comparison of models based on all classifiers for basic vs enhanced features.

# 15 LOFO Comparison: Basic vs Enhanced

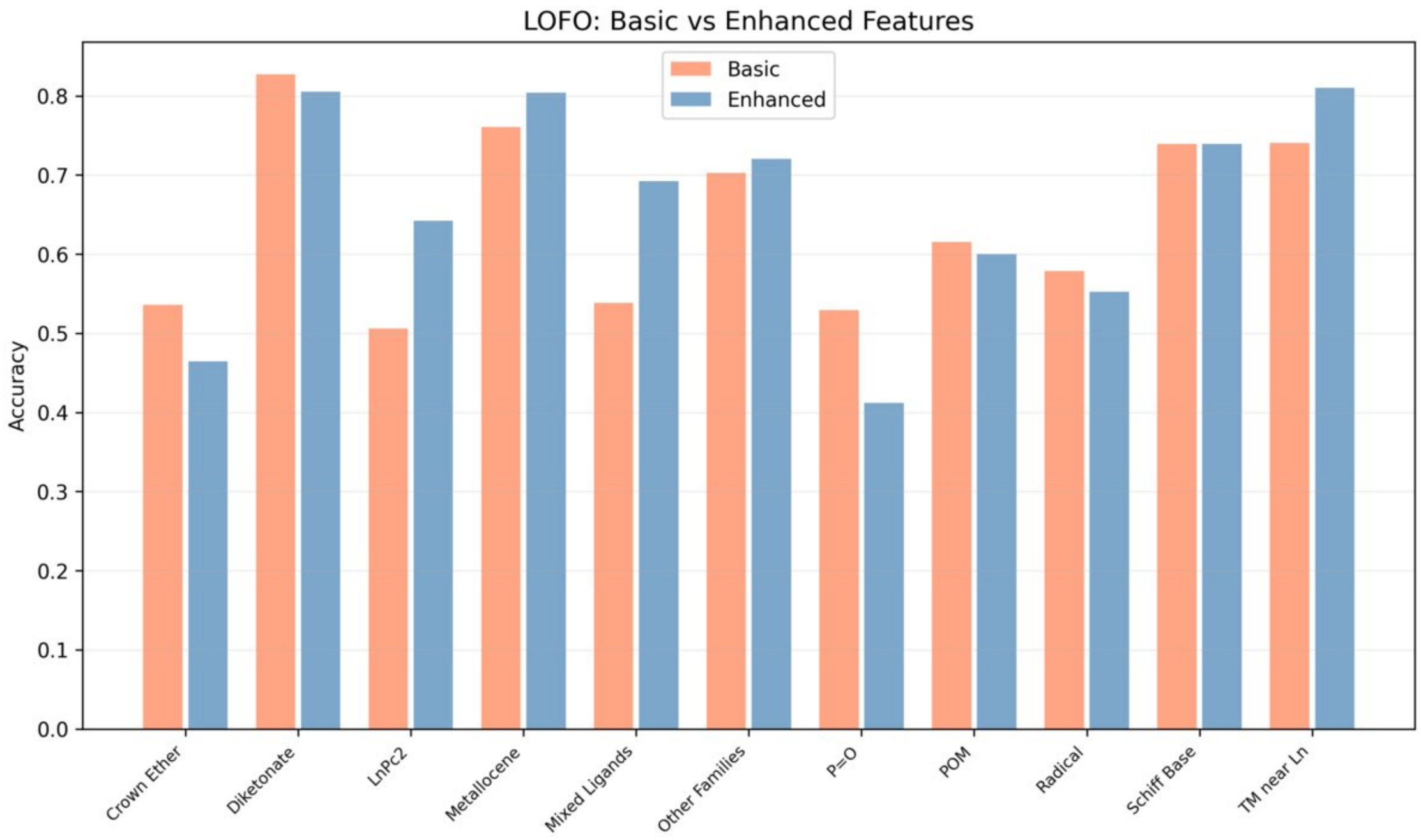


Figure S12: Leave-one-family-out accuracy comparison between basic (30 features) and enhanced (48 features) models.

# 16 Learning Curves

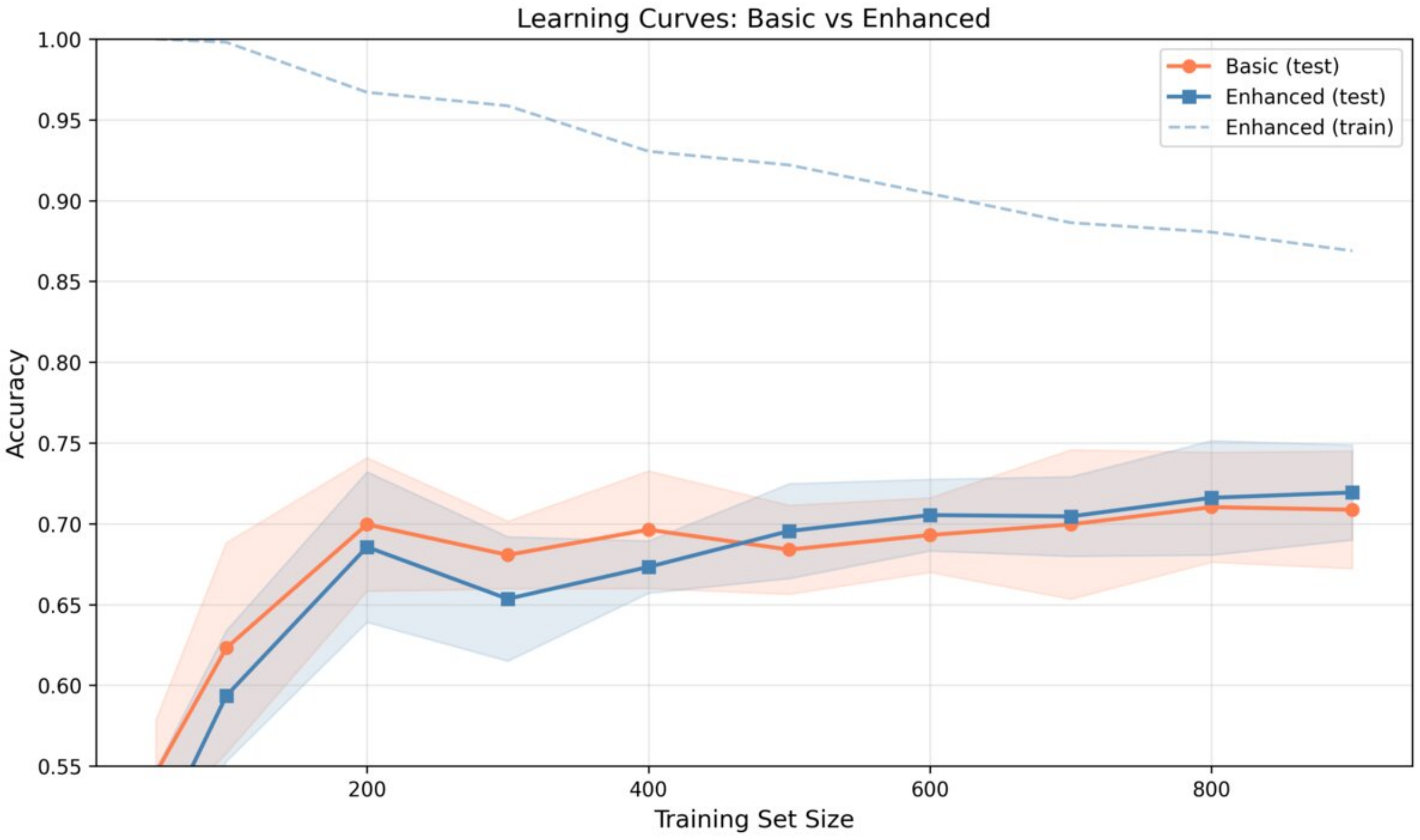


Figure S13: Training and test accuracy curves for both basic and enhanced features under the subsampling protocol. There is still no plateau of the test accuracy at $N = 1000$; there is a greater training–test gap observed in the case of enhanced features.

# 17 SHAP Dependence of CSM Features

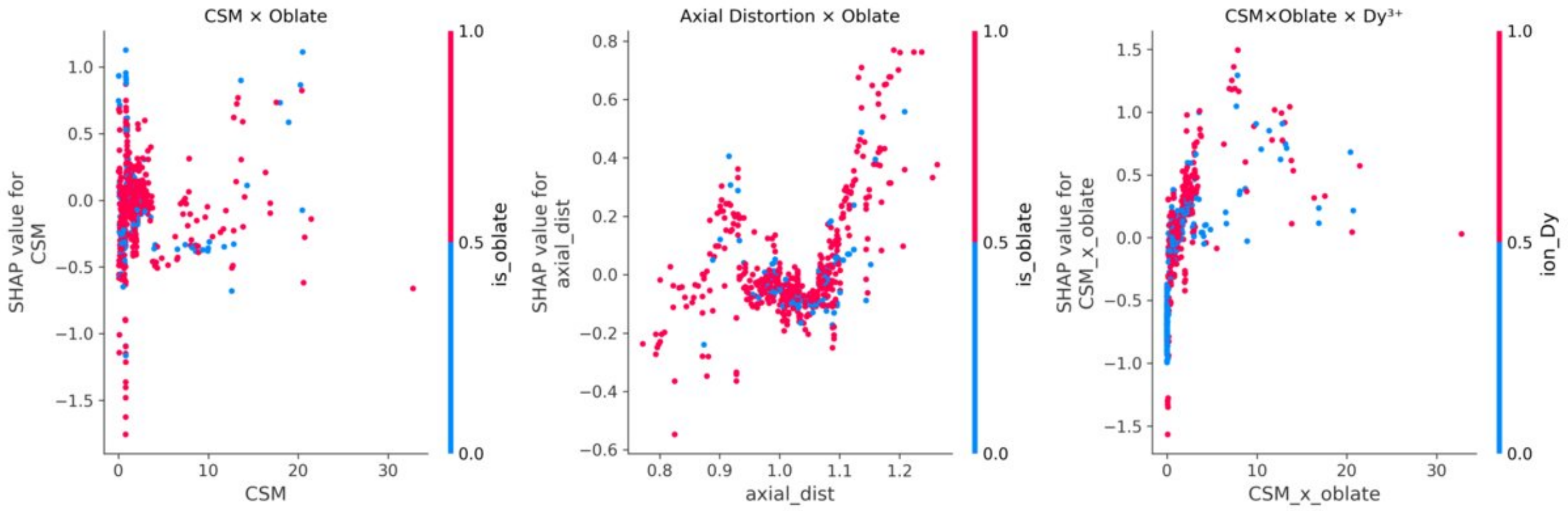


Figure S14: SHAP dependence plots for CSM, axial distortion, and the interaction CSM×oblate, displaying nonlinear and context-dependent effects on the model output.

# 18 Full Post-Hoc Misclassification Analysis

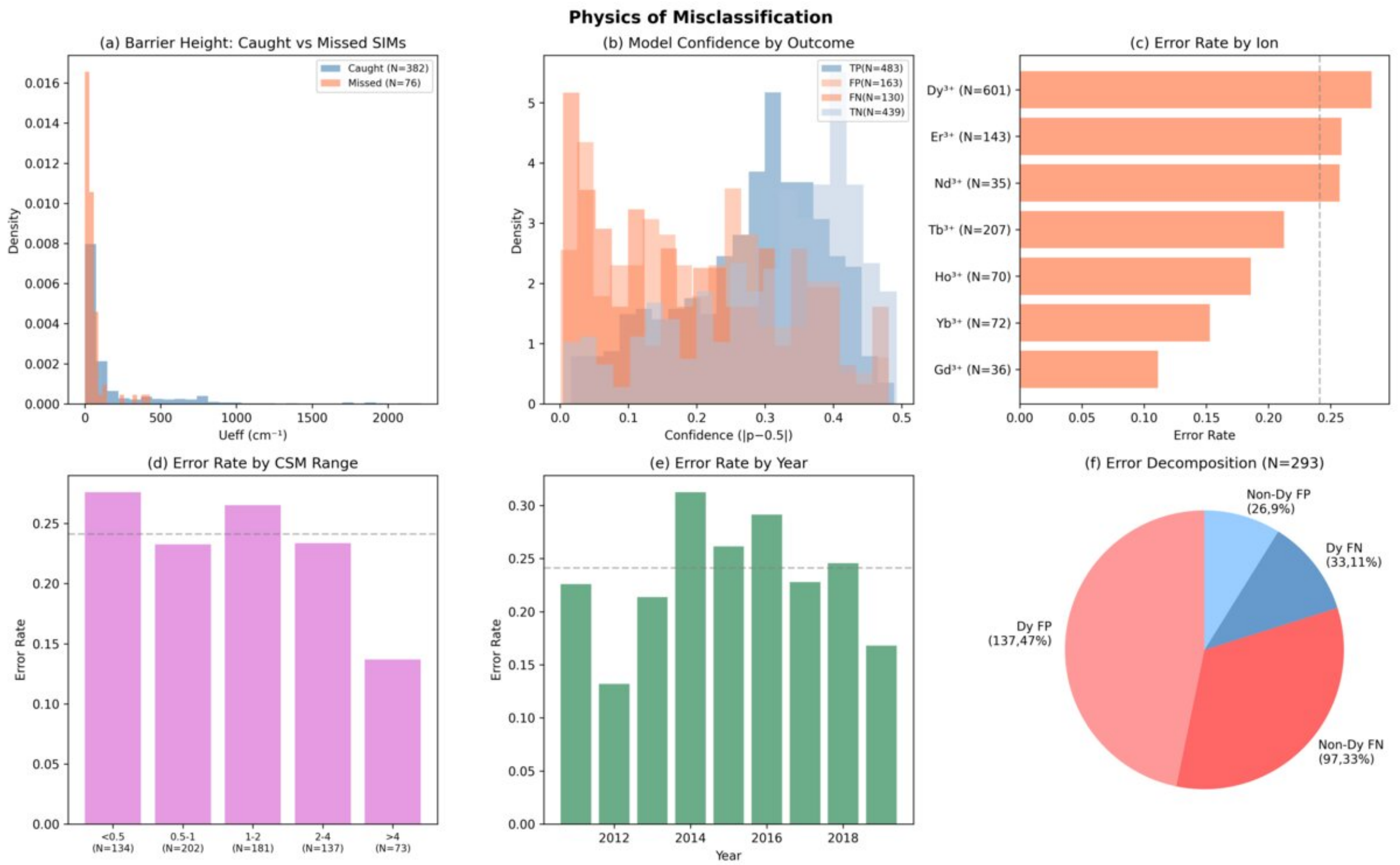


Figure S15: Six-panel full post-hoc analysis of misclassified cases: (a) distribution of $U_{\text{eff}}$ for correctly classified and misclassified SIMs; (b) model confidence per outcome classification; (c) error rates by ion; (d) error rates by CSM range; (e) error rates by publication year; and (f) error decomposition. Panels (a), (c), and (f) are shown as Figure 8 in the main text.

# 19 Point-Charge Crystal-Field Descriptors

To test whether a physically motivated, coordinate-based representation could improve classification accuracy, we derived point-charge crystal-field descriptors from the COD crystal structures. These comprised the Stevens parameters $B_k^q$ and the corresponding $A_k^q$ through rank six, their axial fractions and axiality ratios, and simple electrostatic measures of the ligand charge distribution. The descriptors were successfully constructed for 482 compounds.

Table S3 compares gradient-boosting classifiers trained on the tabular features alone, the 25 crystal-field descriptors alone, and the combined feature set, using the same hyperparameters and five-fold cross-validation procedure as in the main text. The crystal-field descriptors

alone yield an accuracy of 64%. Adding them to the tabular features reduces accuracy from 75.7% to 71.8%, corresponding to a net increase of 19 misclassifications (McNemar's test, $p = 0.08$). A scalar measure of crystal-field anisotropy compatibility derived from the same parameters likewise shows no evidence of separation between the SIM and non-SIM classes (Mann–Whitney test, $p = 0.73$).

An independent implementation based on Wybourne parameters $B_k^q$ ($k = 2, 4, 6$) with Freeman–Watson radial expectation values, applied to all 606 compounds of Section 3.7, gives the same picture (Table S3, lower block): the crystal-field parameters alone yield 62% accuracy (66% for a reduced set of 19 key features), and adding them to the tabular features changes the accuracy by an insignificant amount (+1.2 points, McNemar's test, $p = 0.59$). Compounds whose $B_2^0$ sign matches the anisotropy preference of the ion are SIMs at the same rate as those where it does not (49% vs 54%), and the axiality medians of the SIM and non-SIM classes coincide (0.364 vs 0.368).

Thus, within the present modeling framework, the point-charge crystal-field descriptors provide no improvement over the tabular baseline. This finding is consistent with the interpretation in Section 3.7 of the main text: improving classification may require a more faithful account of the ion's electronic structure in its ligand field, rather than a more detailed description of the geometry alone.

Table S3: Classification performance of point-charge crystal-field (CF) descriptors, tabular features, and their combination. Upper block: Stevens parameters for the same 482 compounds; accuracies are reported as the mean ± standard deviation across five cross-validation folds. Lower block: Wybourne parameters for the 606 compounds of Section 3.7.

| Model | $N_{\text{feat}}$ | Accuracy | AUC |
|---|---|---|---|
| Tabular | 31 | $0.757 \pm 0.022$ | 0.817 |
| CF only | 25 | $0.643 \pm 0.044$ | 0.686 |
| Tabular + CF | 56 | $0.718 \pm 0.042$ | 0.775 |
| Tabular (CF subset) | 30 | 0.756 | 0.817 |
| CF parameters only | 46 | 0.621 | 0.692 |
| CF key features only | 19 | 0.665 | 0.710 |
| Tabular + CF key | 49 | 0.754 | 0.826 |
| Tabular + CF all | 76 | 0.767 | 0.816 |

# 20 Cross-Platform Reproducibility of Ab Initio Calculations

CASSCF/RASSI-SO/SINGLE_ANISO results for COD 4349807 have been calculated independently on two machines with two different versions of the code: OpenMolcas v26.02 (NERSC Perlmutter) and OpenMolcas v26.06 (OLCF Frontier, compiled with GCC 13 against an ILP64 OpenBLAS v0.3.28 build). The $g$ tensor for the ground doublet was reproduced to all reported digits: $(g_x, g_y, g_z) = (0.652, 1.062, 7.177)$.



\end{document}